%% file: Main_Manuscript.tex
\documentclass[twocolumn,twocolappendix]{aastex631}

\usepackage{listings}
\usepackage{multirow}
\usepackage{breqn}
\usepackage{subfigure}
\graphicspath{{./}{figures/}}

\begin{document}

\title{SZ-X-ray Surface Brightness Fluctuations in the SPT-XMM clusters}
\input{authors.tex}

\begin{abstract}

    The hot plasma in galaxy clusters, the intracluster medium (ICM), is expected to be shaped by subsonic turbulent motions, which are key for heating, cooling, and transport mechanisms. The turbulent motions contribute to the non-thermal pressure which, if not accounted for, consequently imparts a hydrostatic mass bias. Accessing information about turbulent motions is thus of major astrophysical and cosmological interest. Characteristics of turbulent motions can be indirectly accessed through surface brightness fluctuations. 
    This study expands on our pilot investigations of surface brightness fluctuations in the SZ and X-ray by examining, for the first time, a large sample of 60 clusters using \textit{both} SPT-SZ and \textit{XMM-Newton} data and span the redshift range $0.2 < z < 1.5$,
     thus constraining the respective pressure and density fluctuations within 0.6~$R_{500}$. We deem density fluctuations to be of sufficient quality for 32 clusters, finding mild correlations between the peak of the amplitude spectra of density fluctuations and various dynamical parameters.    
    We infer turbulent velocities from density fluctuations with an average Mach number $\mathcal{M}_{\text{3D}} = 0.52 \pm 0.14$, in agreement with numerical simulations. For clusters with inferred turbulent Mach numbers from both pressure, $\mathcal{M}_{\text{P}}$ and density fluctuations, $\mathcal{M}_{\rho}$, we find broad agreement between $\mathcal{M}_{\text{P}}$ and $\mathcal{M}_{\rho}$. Our results suggest either a bimodal or skewed unimodal Mach number distribution, with the majority of clusters being turbulence-dominated (subsonic) while the remainder are shock-dominated (supersonic).
    
    %We infer that accessing density and pressure fluctuations out to $R_{500}$ across the mass and redshift range probed here are beyond the practical capabilities of current facilities.
    %With the advancement of Sunyaev-Zel'dovich (SZ) studies and surveys relative to X-ray observations, we seek to investigate surface brightness fluctuations in a sample of SPT-SZ clusters which also have archival \textit{XMM-Newton} data. 

\end{abstract}


\section{Introduction} \label{sec:intro} \input{Introduction}

\section{Approach} \label{sec:approach} \input{Approach}

\section{Results} \label{sec:results} \input{Results}

\section{Discussion} \label{sec:discussion} \input{Discussion}

\section{Conclusions} \label{sec:conclusions} \input{Conclusions}

\begin{acknowledgments}
\textit{Acknowledgements.}
The authors would like to thank the anonymous referee for comments which have improved this work.

CR acknowledges support from NASA ADAP grant 80NSSC19K0574 and Chandra grant G08-19117X. 
MG acknowledges funding support from the ERC Consolidator Grant \textit{BlackHoleWeather} (101086804). RK acknowledges support from the Smithsonian Institution, the Chandra High Resolution Camera Project through NASA contract NAS8-03060, and NASA Grants 80NSSC19K0116, GO1-22132X, and GO9-20109X. PN was supported by NASA contract NAS8-03060. CLR acknowledges support from the Australian Research Council’s Discovery Project scheme (No. DP200101068). YS acknowledges support from Chandra grants GO1-22126X and GO2-23120X. 
EB acknowledges financial support from the ERC Consolidator Grant \textit{DarkQuest} (101002585). WF acknowledges support from the Smithsonian Institution, the Chandra High Resolution Camera Project through NASA contract NAS8-0306, NASA Grant 80NSSC19K0116 and Chandra Grant GO1-22132X. 

The South Pole Telescope program is supported by the National Science Foundation (NSF) through awards OPP-1852617 and OPP-2332483. Partial support is also provided by the Kavli Institute of Cosmological Physics at the University of Chicago. Argonne National Laboratory's work was supported by the U.S. Department of Energy, Office of High Energy Physics, under contract DE-AC02-06CH11357. Work at Fermi National Accelerator Laboratory, a DOE-OS, HEP User Facility managed by the Fermi Research Alliance, LLC, was supported under Contract No. DE-AC02-07CH11359.
\end{acknowledgments}

\vspace{5mm}
\facilities{SPT, \textit{XMM-Newton}}

\software{astropy \citep{astropy2013,astropy2018,astropy2022},  emcee \citep{foreman2013}, pyproffit\citep{eckert2017}, ESAS \citep{snowden2008} }



\bibliography{July2024_References}{}
\bibliographystyle{aasjournal}



\appendix

\section{Cluster properties} \label{sec:app_table}
\input{Appendix_Table}

\section{Masking Substructure} \label{sec:substructure_masking}
\input{Appendix_Substructure}

\section{Amplitude Spectra} \label{sec:app_spectra}
\input{Appendix_Spectra}

%\section{Additional correlations} \label{sec:app_AddCorr}
%\input{Appendix_FurtherCorrelations}

\section{Constraints out to $R_{500}$} \label{sec:app_ring2}
\input{Appendix_Ring2}

%\section{Respository of Figures} \label{sec:Repository}
%\input{RepositoryOfFigures}

%\newpage
%\section{Correlations with different significance thresholds} \label{sec:ToBeDeleted}
%\input{CompareCorrelations}

%\newpage
%\section{Correlations with different Mach upper limits} \label{sec:ToBeDeleted2}
%\input{CompareCorrelations2}

\end{document}

%% file: authors.tex
\author[0000-0001-5725-0359]{Charles E. Romero}
\altaffiliation{E-mail: \href{mailto:charles.romero@gmail.com}{charles.romero@gmail.com} }
\affiliation{Center for Astrophysics $\vert$ Harvard \& Smithsonian, 60 Garden Street, Cambridge, MA 02138, USA}

\author[0000-0003-2754-9258]{Massimo Gaspari}
\affiliation{Department of Physics, Informatics \& Mathematics, University of Modena \& Reggio Emilia, 41125 MO, IT}

\author[0000-0002-4962-0740]{Gerrit Schellenberger}
\affiliation{Center for Astrophysics $\vert$ Harvard \& Smithsonian, 60 Garden Street, Cambridge, MA 02138, USA}

\author[0000-0002-5108-6823]{Bradford A. Benson}
\affiliation{Fermi National Accelerator Laboratory, MS209, P.O. Box 500, Batavia, IL 60510, USA}
\affiliation{Department of Astronomy and Astrophysics, University of Chicago, 5640 South Ellis Avenue, Chicago IL 60637, USA}
\affiliation{Kavli Institute for Cosmological Physics, University of Chicago, 5640 South Ellis Avenue, Chicago, IL 60637, USA}

\author[0000-0001-7665-5079]{Lindsey E. Bleem}
\affiliation{High Energy Physics Division, Argonne National Laboratory, 9700 South Cass Avenue, Lemont, IL, 60439, USA}
\affiliation{Kavli Institute for Cosmological Physics, University of Chicago, 5640 South Ellis Avenue, Chicago, IL 60637, USA}

\author[0000-0002-7619-5399]{Esra Bulbul}
\affiliation{Max Planck Institute for Extraterrestrial Physics, Giessenbachstrasse 1, 85748 Garching, Germany}

%\author[0000-0002-8248-4488]{Matthias Klein}
%\affiliation{University Observatory, Faculty of Physics, Ludwig-Maximilians-Universit\"{a}t, Scheinerstr. 1, 81679 Munich, Germany}

\author[0000-0002-9478-1682]{William Forman}
\affiliation{Center for Astrophysics $\vert$ Harvard \& Smithsonian, 60 Garden Street, Cambridge, MA 02138, USA}

\author[0000-0002-0765-0511]{Ralph Kraft}
\affiliation{Center for Astrophysics $\vert$ Harvard \& Smithsonian, 60 Garden Street, Cambridge, MA 02138, USA}

\author[0000-0003-0297-4493]{Paul Nulsen}
%\altaffiliation{ICRAR, University of Western Australia, 35 Stirling Hwy, Crawley, WA 6009, Australia}
\affiliation{Center for Astrophysics $\vert$ Harvard \& Smithsonian, 60 Garden Street, Cambridge, MA 02138, USA}
\affiliation{ICRAR, University of Western Australia, 35 Stirling Hwy, Crawley, WA 6009, Australia}

\author[0000-0003-2226-9169]{Christian L. Reichardt}
\affiliation{School of Physics, University of Melbourne, Parkville, VIC 3010, Australia}

%\author{Laura Salvati}
%\affiliation{Universit\'{e} Paris-Saclay, CNRS, Institut d'Astrophysique Spatial, F-91405, Orsay, France}

\author[0000-0002-5222-1337]{Arnab Sarkar}
\affiliation{Kavli Institute for Astrophysics and Space Research, Massachusetts Institute of Technology, 70 Vassar Street, Cambridge, MA 02139, USA}

\author[0000-0003-3521-3631]{Taweewat Somboonpanyakul}
%\affiliation{Kavli Institute for Particle Astrophysics and Cosmology, Stanford University, 452 Lomital Mall, Stanford, CA 94305, USA}
\affiliation{Department of Physics, Faculty of Science, Chulalongkorn University
254 Phayathai Road, Pathumwan, Bangkok Thailand. 10330}

\author[0000-0002-3886-1258]{Yuanyuan Su}
\affiliation{Department of Physics and Astronomy, University of Kentucky, 505 Rose Street, Lexington, KY 40506, USA}

%\affiliation{Various}

%% file: Introduction.tex
    The dominant baryonic component of galaxy clusters is the hot ($10^7$ to $10^8$ K) intracluster medium (ICM). The thermal component of the ICM is observable via X-rays and the millimeter band via the Sunyaev-Zel'dovich (SZ) effect \citep{sunyaev1972}, while relativistic particles are observable via synchrotron radiation at lower frequencies. The thermal gas, especially at moderate to large radii ($\sim$\,$R_{2500}$ to $R_{500}$)\footnote{For a density contrast, $\Delta$, $R_{\Delta}$ is the radius within which the mean matter density is $\Delta$ times the critical density of the universe.} matches well expectations of self-similarity and gravitational heating \citep[see e.g.][]{kravtsov2012}. Gravitational heating is likely to proceed primarily through shock (adiabatic) heating or turbulent (dissipative) heating. Through numerical simulations \citet{shi2020MNRAS} found that turbulent heating should be dominant within $R_{500}$, while shock heating (especially for accretion shocks) will dominate at $r > R_{500}$. 

    In the central regions of galaxy clusters, baryonic physics is critical. In particular, radiative cooling appears to be self-regulated via active galactic nuclei (AGN) feedback \citep[e.g][]{mcnamara2012,Gaspari2014_AGN,voit2017}. In the case of merger shocks and AGN feedback, much of the gas heating will be localized and yet the balanced feedback requires much of the energy to be isotropically distributed throughout the ICM, with turbulence likely playing a crucial role in this transport \citep[e.g.][]{wittor2020,wittor2023}. At the same time, turbulence is expected to have a prominent role in gas condensation in cluster centers driving chaotic cold accretion onto supermassive black holes (\citealt{gaspari2020}, for a review), as well as in the re-acceleration of cosmic rays generating extended radio emission \citep{BrunettiJones2014,eckert2017,vanweeren2019,Pasini2024}.

    Beyond the inextricable role that turbulence plays in the thermodynamics of the ICM, it will also impart a non-thermal pressure component throughout the ICM and thereby contribute to the hydrostatic mass bias\footnote{The hydrostatic mass bias is the bias on a mass estimate when assuming that the cluster is in hydrostatic equilibrium and thus only supported by thermal pressure.}, which is currently the dominant source of systematic uncertainty in mass estimation from the ICM \citep[see][for a recent review]{Pratt2019}.

    Despite the integral role that turbulence must play in the observed X-ray, millimeter, and radio signals, constraining turbulence, especially turbulent motions has not come easily \citep[e.g.][]{simionescu2019}. While there is much anticipation of turbulent velocity constraints from Doppler shifts and broadening of lines in high-resolution X-ray spectroscopy via the recently launched \textit{XRISM} \citep{xrism2020} and proposed facilities such as \textit{LEM} \citep{kraft2022}, \textit{Athena} \citep{athena2013,rau2013,meidinger2017}, and now \textit{NewAthena} \citep{NewAthena2025NatAs} turbulent velocities can also be accessed via the driven surface brightness fluctuations in the X-ray \citep[e.g.][]{schuecker2004,churazov2012,Gaspari2013_PS,Gaspari2014_PS,hofmann2016,heinrich2024} and SZ images \citep[][]{khatri2016,romero2023}. From a cosmological perspective, constraining turbulent motions at cluster outskirts (at radii of roughly $R_{500}$ and larger) is most important. As both SZ and X-ray signals are faint (relative to the cluster cores), such observations are expensive. Moreover, the relative expense of velocity constraints from high-resolution X-ray spectroscopy compared to surface brightness fluctuations \citep[e.g.][]{romero2024b}, suggests that measuring turbulence in the cluster outskirts is more feasible via surface brightness fluctuations.

    %\textcolor{red}{I want to mention something about density versus pressure fluctuations here. Also the Mach relation(s).}
    Accessing surface brightness fluctuations via both X-ray and SZ can be desirable for their different observational properties, i.e. advantages, as well as their different physical insights, where they are best suited to constraining density and pressure fluctuations, respectively. With both density and pressure fluctuations, one can assess the effective equation of state \citep{romero2023} and potentially more robustly constrain the turbulent gas velocities.

    Given the required sensitivity to obtain meaningful constraints on density and pressure fluctuations, we may also be interested in correlating those fluctuations with other, more accessible parameters. While AGN feedback should generate turbulence in the central regions (e.g.~\citealt{wittor2023}), this is not expected to be dominant at moderate ($r > R_{2500c}$) cluster-centric radii (e.g.~\citealt{lau2017}). If merger activity is expected to be the primary driver of gas motion at moderate radii and larger, then we can expect some degree of correlation with typical dynamical parameters (e.g.~\citealt{lovisari2017,yuan2022}).
    
    %%%%%%%%%%%%%%%%%%%%%%%%%%%%%%%%%%%%%%%%%%%%%%%%%%%%%%%%%%%%%%%%%%%%%%%%%%%%%%%%%%%%%%%%%%%%%%%%%%%%%%%%%%%%%%%%%%%%%%%%%%%%%%%%%%
    %%% %%% %%% %%%                   TRANSITION FROM BACKGROUND TO A QUICK OVERVIEW OF THIS STUDY                    %%% %%% %%%  %%% 
    %%%%%%%%%%%%%%%%%%%%%%%%%%%%%%%%%%%%%%%%%%%%%%%%%%%%%%%%%%%%%%%%%%%%%%%%%%%%%%%%%%%%%%%%%%%%%%%%%%%%%%%%%%%%%%%%%%%%%%%%%%%%%%%%%%
    %%% (PICKUP HERE. 09 August)

    %In this work we complete our analysis across a sample of clusters which have sufficient SPT and \textit{XMM-Newton} data \citep{bulbul2019}.
    The sample selection and approach, building on the pilot study of \citet{romero2024} are discussed in Section~\ref{sec:approach}. We present results in Section~\ref{sec:results} and discuss them in Section~\ref{sec:discussion}.
    Our assumed cosmology adopts $H_0 = 70$~km~s$^{-1}$~Mpc$^{-1}$, $\Omega_M = 0.3$, $\Omega_{\Lambda} = 0.7$. We report all uncertainties as one standard deviation (for distributions taken to be symmetric) or the distance from the median to the 16th and 84th percentiles (when allowing for asymmetric distributions), unless otherwise stated.

%% file: Approach.tex
    %\textcolor{red}{Write an overview paragraph before subsection.}
    %\subsection{Sample Selection} \label{sec:sample}

        \begin{figure}
            \centering
            \includegraphics[width=0.47\textwidth]{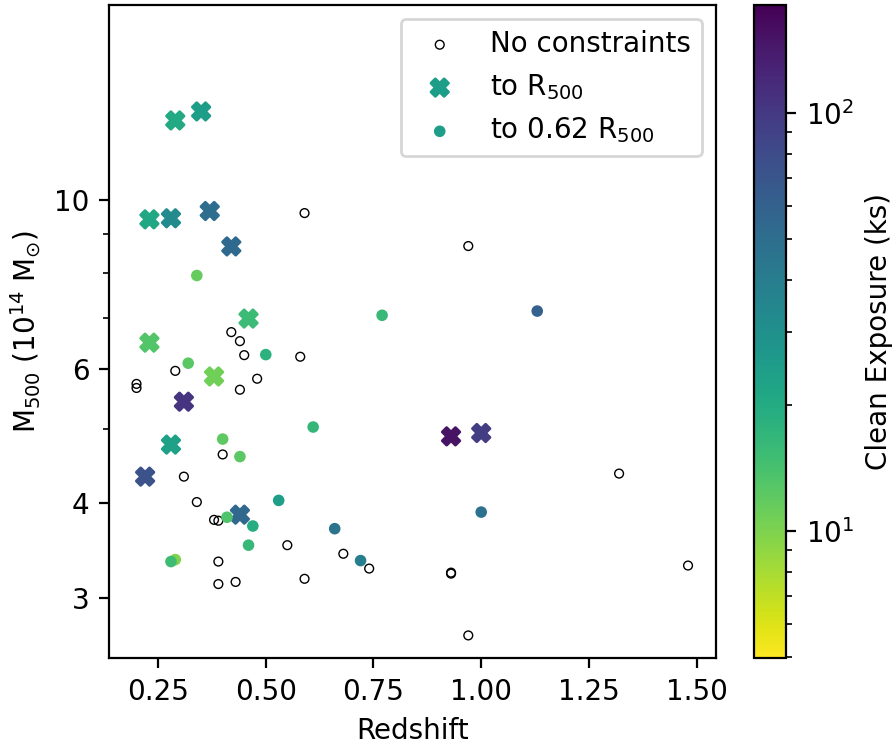}
            \caption{The mass and redshift distribution of clusters in our {\bf (SPT-XMM)} sample as well as information regarding constraints on surface brightness fluctuations from \textit{XMM-Newton} data. Empty circles denote no significant constraints are reported; filled markers denote that constraints of at least $2\sigma$ were obtained within $0.62 R_{500}$ (filled circles) and out to $R_{500}$ (crosses). The color indicates a representative exposure time across the EPIC cameras.}
            \label{fig:sample_dist}
        \end{figure}

        We seek to constrain density and pressure fluctuations via X-ray and SZ observations, respectively. Unlike previous studies, we aim to have a large statistical sample across a wide mass and redshift range (see Figure~\ref{fig:sample_dist}). To do this, we use a sample defined by the galaxy clusters in the SPT-SZ survey \citep{bleem2015} which also have sufficiently deep \textit{XMM-Newton} data. Such a sample was compiled in \citet{bulbul2019}.
        %they note that the median redshift of their sample is 0.45, which is lower than the median redshift ($z = 0.55$) used in the cosmology sample of \citet{deHaan2016}. The median mass ($M_{500}$) of the sample in \citet{bulbul2019} is $4.77 \times 10^{14} \text{M}_{\odot}$, compared to a median mass $M_{500} = 4.57 \times 10^{14}  \text{M}_{\odot}$ in \citet{deHaan2016}. 
        SPT-CLJ0014-3022 was observed with the SPT camera \citep{plagge2010} separate from the SPT-SZ survey. Given that SPT-CLJ0014-3022 (also called Abell 2744) meets the redshift ($z > 0.2$) and photon count (more than 1000 filtered source counts in MOS cameras) criteria for the sample selection in \citep{bulbul2019}, it is also included in our sample, which we call the SPT-XMM sample. Additional properties of the sample are discussed in Appendix~\ref{sec:app_table}.

    \subsection{Image and Fourier analysis}
    
        Our data analysis approach follows that used in our precursory studies (\citealt{romero2024,khatri2016}), which we summarize here.      
        For surface brightness images, $y$ and $S$, in the SZ and X-ray, respectively, we fit smooth surface brightness models, $\bar{y}$ and $\bar{S}$, to their respective images. In this work, we take our models to be circular $\beta$ models in both the SZ and X-ray cases, with the SZ and X-ray centers fixed to the centroid found in the X-ray dataset. We also run the SZ profile fitting procedure with the center free to infer $\Delta_c$, the SZ to X-ray centroid offset.
        The SZ model, $\bar{y}$, is entirely defined by the ICM; i.e. any mean level or background component is assumed to have been nulled. The X-ray model, $\bar{S}$, can be taken as the sum of an ICM component and a background component: $\bar{S} = \bar{S}_{\text{ICM}} + \bar{S}_{\text{bkg}}$ \citep{romero2023}. Residual maps are taken to be $\delta y = y - \bar{y}$ and $\delta S = S - \bar{S}$. Point sources and chip gaps are masked as in previous analyses \citep{romero2023,romero2024}.
        
        We characterize surface brightness fluctuations of the normalized residual maps, $\delta y / \bar{y}$ and $\delta S / \bar{S}_{\text{ICM}}$, via a wavelet decomposition method based on a Mexican Hat filter (\citealt{arevalo2012}). As in \citet{romero2024}, we adhere to exploring fluctuations in two regions: Ring 1 being a circle of radius $0.62 R_{500}$ and Ring 2 being the annulus between $0.62 R_{500}$ and $R_{500}$. 

        The power spectra of surface brightness fluctuations in SZ and X-ray are then deprojected to pressure and density fluctuations, characterized by their 3D spectra, $P_{\text{3D}}$, as in \citet{romero2024}. The fluctuations may also be represented through their amplitude spectra:
        \begin{equation}   
            A_{\text{3D}} = \sqrt{4 \pi k^3 P_{\text{3D}}}. \label{eqn:a3d}
        \end{equation}
        Specifically, we calculate $A_{\text{3D}}$ to correspond to density fluctuations, i.e. $A_{\rho}$, when considering X-ray data. Similarly, in the case of SZ data, $A_{\text{3D}}$ is taken as $A_{\text{P}}$. 
        We sample our spectra at angular scales between our resolution limit (taken to be $10^{\prime\prime}$ for \textit{XMM-Newton} and $1^{\prime}.25$ for SPT) and $\theta_{500}$ (the angular extent of $R_{500}$ on the sky) with logarithmic spacing close to a factor of 2 so that each point is (approximately) independent.
        %for corresponding to the 3D power spectra ($P_{\text{3D}}$) of density or pressure fluctuations. 

        \subsubsection{X-ray image processing and spectral co-addition.}

            We extract images in the [0.4-1.25] keV and [2.0-5.0] keV bands for each of the EPIC cameras through the use of ESAS \citep{snowden2008}, for each ObsID. As in our pilot study, a single cluster center and point source mask is adopted across all images of a particular cluster. A $\beta$ model is fit to each image, and fluctuation (normalized residual) images $S/\bar{S}_{\text{ICM}}$ are produced (for each band, camera, and ObsID). Power spectra are measured on each image following the Delta Variance method employed in \citet{arevalo2012}, and deprojected to power spectra of gas density as in our pilot study \citep{romero2024}. These deprojected power spectra (per band, camera, and ObsID) are combined by taking the weighted average, for a given cluster.
            
            In our pilot study, neither of the two clusters investigated had clear substructure in the \textit{XMM-Newton} images, and we did not investigate masking substructure. In the full sample, we encountered SPT-CLJ0658-5556 (aka the Bullet cluster), SPT-CLJ0304-4401, SPT-CLJ2023-5535, SPT-CLJ0014-3022, and SPT-CLJ0225-4155 which we identified as having significant substructure and mask the substructure according to an algorithm detailed in Appendix~\ref{sec:substructure_masking}. 
        
        \subsubsection{SZ analysis of SPT-SZ clusters} 

            Our analyses of SPT images proceed as in \citet{romero2024} with the exception of the analysis of SPT-CLJ0014-3022 which is not in the SPT-SZ survey. SPT images are taken as minimum-variance Compton $y$ maps \citep{bleem2022}.

        \subsubsection{SZ analysis of SPT-SZ clusters SPT-CLJ0014-3022}

            The dataset for SPT-CLJ0014-3022 \citep{Crawford2022}
            %\footnote{Available at \url{https://doi.org/10.17038/HEP/2342125}.}
            does not include half maps, but rather a single map out to large cluster-centric radii. The map is tapered starting at roughly $3 R_{500}$. Two point sources are evident in the map (both are beyond $2 R_{500}$) and are masked. 
            %\footenote{Available at \href{https://doi.org/10.17038/HEP/2342125}{https://doi.org/10.17038/HEP/2342125} }
            
            A mean level is found at radii beyond $R_{500}$ and is subtracted. A $\beta$ model is then fit to the cluster and a power spectrum within Ring 1 can be calculated on the resultant $\delta y/\bar{y}$ image, which will include power from the noise. Power spectra are computed in six non-overlapping regions of equivalent radius of Ring 1 spaced far from the cluster center. That is, for each region, $i$, a corresponding $(\delta y/\bar{y})_i$ map is computed by moving the $\bar{y}$ model center to the center of the region. From these power spectra of ``noise realizations", we debias and derive uncertainties on the desired SZ surface brightness fluctuations.

            %\textcolor{red}{No substructure masking for SPT images.}

    %\subsection{Correlations}

%% file: Results.tex
    %%% This figure introduced in Section 3.1
    \begin{figure*}[!ht]
        \centering
        \subfigure{\includegraphics[width=0.7\textwidth]{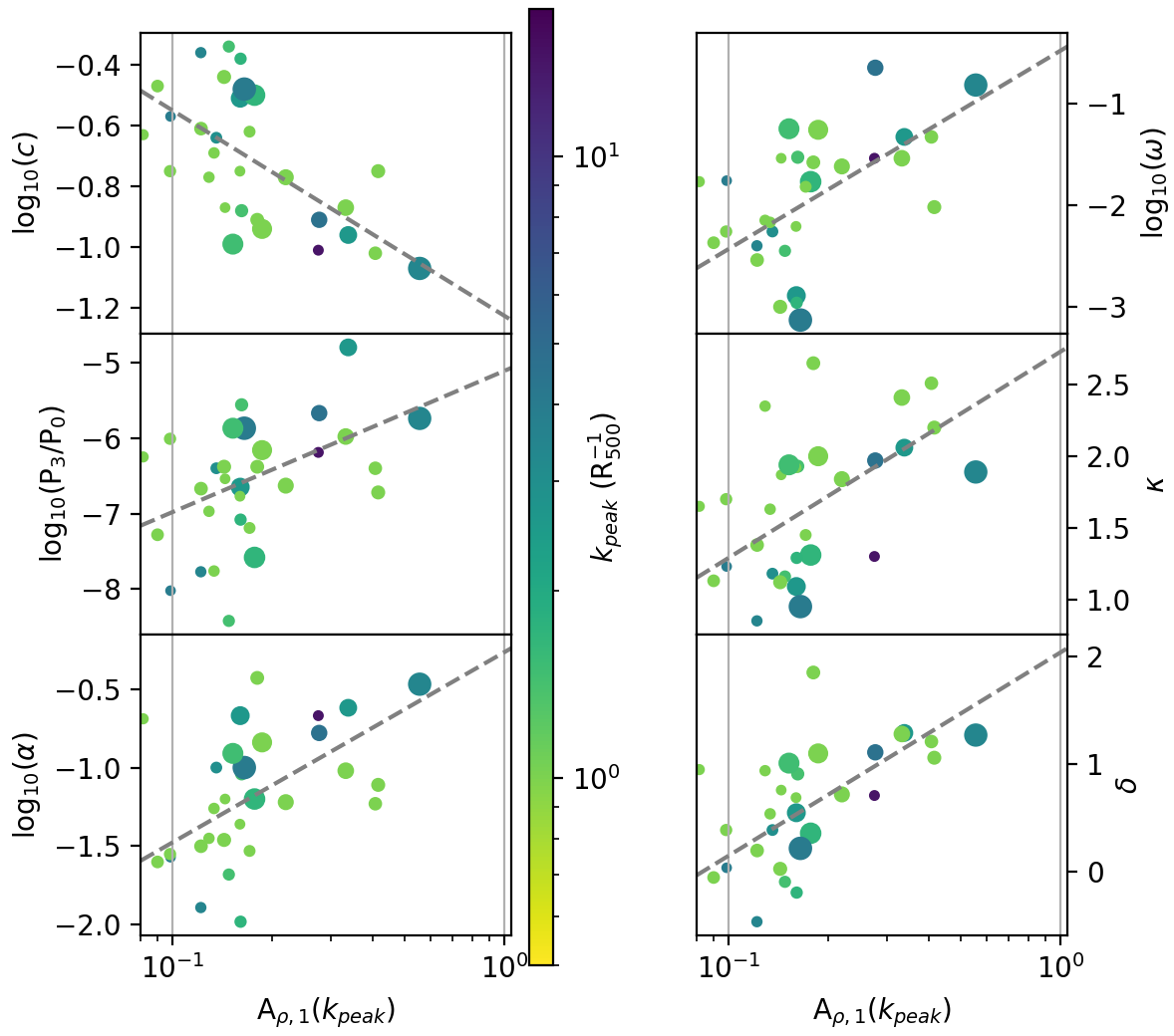}}
        \caption{Scatter plots of dynamical parameters relative to the inferred $A_{\rho}$ in Ring 1. Sizes of markers indicate the maximal $\xi$ (see Appendix~\ref{sec:app_table}) in $A_{\rho,1}$. The coloring of the markers corresponds to the location (wavenumber) of the inferred peak, adopting a SNR threshold of $\xi > 2$.
        }
        \label{fig:correlation1}
    \end{figure*}
    
    \begin{figure*}[!ht]
        \centering
        \subfigure{\includegraphics[width=0.7\textwidth]{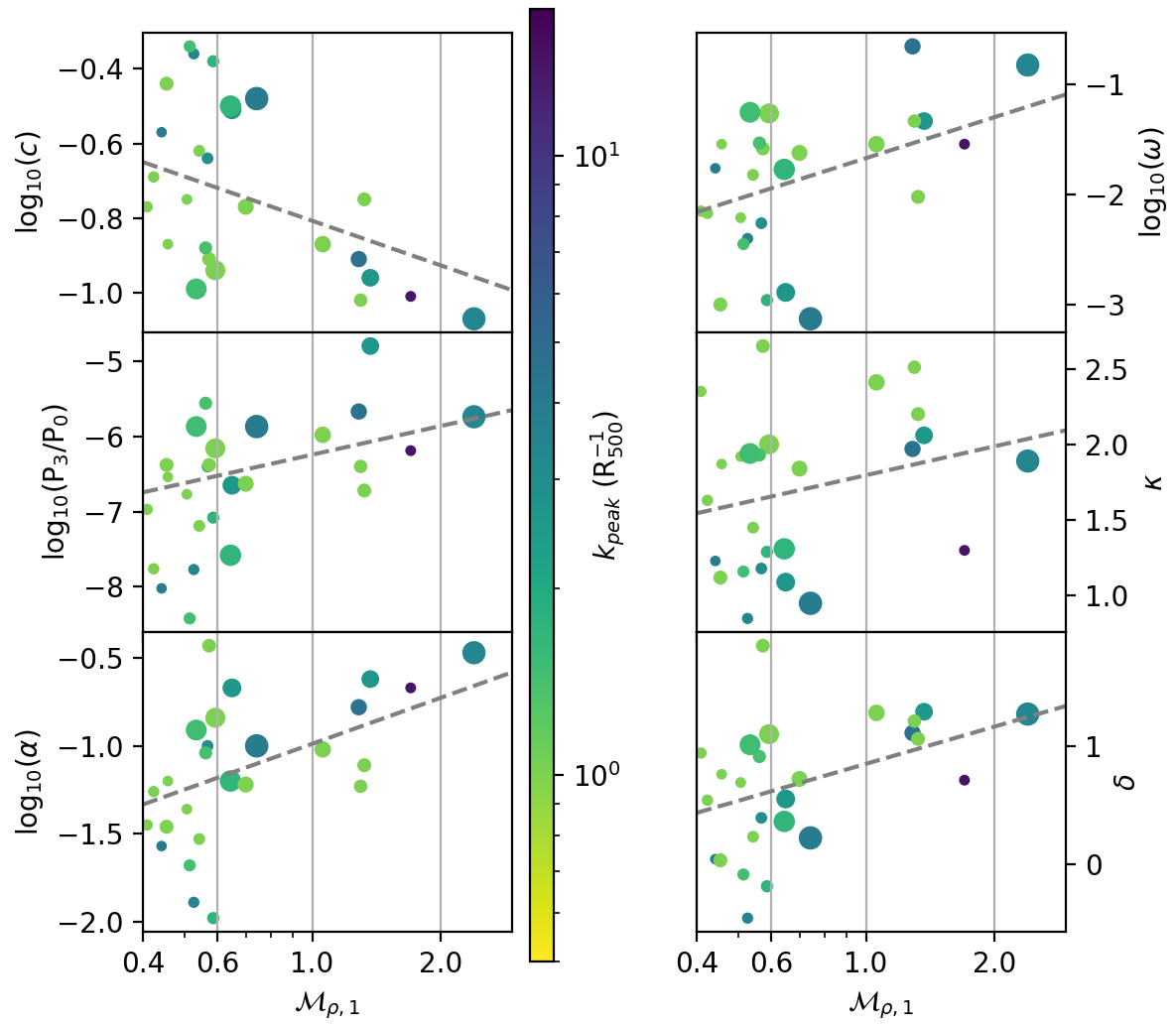}}
        \caption{As with Figure~\ref{fig:correlation1}, but relative to the inferred turbulent 3D Mach number $\mathcal{M}_{\rho}$ in Ring 1.
        }
        \label{fig:correlation2}
    \end{figure*}

    From pressure or density fluctuations, one can infer turbulent velocities quantified in relation to the sound speed, i.e.~their Mach numbers \citep[e.g][]{Gaspari2014_PS,khatri2016,romero2023,dupourque2023,heinrich2024}. In particular, one either integrates over the power spectrum \citep[][]{simonte2022,zhuravleva2023} and applies a linear relation to obtain a Mach number, or one finds the peak of the amplitude spectrum and applies a linear scaling from that peak to obtain a Mach number \citep{Gaspari2013_PS}. As we do not always have good constraints at all scales of our power spectra, we opt to estimate Mach numbers from our defined peak of each amplitude spectrum.
    It is interesting to note that such a linear relation might appear a trivial result (first shown in \citealt{Gaspari2013_PS}), however, this linearity only arises in stratified atmospheres (like the ICM), while a quadratic scaling is expected in pure hydrodynamics (\citealt{churazov2012}).

    Throughout this paper, we present Mach numbers in terms of 3D gas velocities, i.e.~$\mathcal{M}_{\text{3D}}$. When these Mach numbers are (specifically) inferred from density and pressure fluctuations, we adopt the respective notations $\mathcal{M}_{\rho}$ and $\mathcal{M}_{P}$. At times we further specify which rings these values may pertain to with an additional index (subscript), e.g.~for Ring 1:  $\mathcal{M}_{\rho,1}$ and $\mathcal{M}_{P,1}$. To determine turbulent velocities, we adopt the relations from \citet{Gaspari2013_PS}:
    \begin{align}
        \mathcal{M}_{\rho} &= 4.0 A_{\rho}(k_{\text{peak},\rho}) \left( \frac{l_{\text{inj}}}{0.4 R_{500}} \right)^{\alpha_{\rm H}} \label{eqn:MachXR} \\
        \mathcal{M}_{P} &= 2.4 A_{P}(k_{\text{peak},P})  \left( \frac{l_{\text{inj}}}{0.4 R_{500}} \right)^{\alpha_{\rm H}}, \label{eqn:MachSZ}
    \end{align}
    where $l_{\text{inj}}$ is the injection scale and the parameter $\alpha_{\rm H}=-0.25$ models the hydrodynamical regime of negligible thermal conduction, as expected in the ICM due to magnetic and plasma micro-scale processes (\citealt{Gaspari2014_PS,ZuHone2015,Komarov2016}).

    We define the peak of $A_{\rho}$ to be the maximum of the set of points with signal-to-noise ratio (SNR), $\xi_{A_{\rho}} = A_{\rho} / \sigma_{A_{\rho}}$, greater than 2 (amplitude spectra with at least one node $\xi > 9$ are shown in Appendix ~\ref{sec:app_spectra}). By extension, we define $k_{\text{peak}}$ to be the wavenumber at which this peak is found. For a well-sampled and well-constrained amplitude spectrum, the inverse of the injection scale, $k_{\text{inj}} = 1/l_{\text{inj}}$, will effectively be the same as $k_{\text{peak}}$. However, our spectra are not well sampled, and thus we simply take $k_{\text{peak}}$ as a proxy for $k_{\text{inj}}$. 

    Of the 60 clusters in our sample, 32 clusters yielded amplitude spectra of density fluctuations where a peak (as defined above) could be identified in Ring 1 and 15 clusters where such a peak can be identified in Ring 2. From the SZ side, only seven clusters are found to have a node in the amplitude spectra of Ring 1 above $1.5 \sigma$. In the following sections we focus on the results within Ring 1.

    %$A_{\rho}(k_{\text{peak}})$

    \subsection{Correlations with dynamical parameters}

        As we may expect the inferred density fluctuations to be related to merger activity, we investigate correlations between the peak of the amplitude spectra of the inner rings (Ring 1) and the dynamic parameters as calculated by \citet{yuan2022}. 
        These parameters are $c$, $P_{3}/P_{0}$, $\alpha$, $\omega$, $\kappa$, and $\delta$ which correspond to a concentration index,  power ratio, asymmetry factor, peak-centroid offset, profile parameter, and morphology index, respectively. The quantitative formulae for these values can be found in \citet{yuan2022}. We take the values published in their table; some of the values are published as the base-10 logarithm of the above parameters, in which case we retain this logarithm.
        
        Figures~\ref{fig:correlation1}-\ref{fig:correlation2} show the retrieved correlations between the dynamical parameters and $A_{\rho}$ or $\mathcal{M}_{\rho}$, respectively. The size of a marker in these figures corresponds to that maximal significance, $\xi$, in the amplitude spectrum for a given cluster. While the peak itself may be less significant, this is a means of indicating the overall quality of the data. The color of the points also indicates the inferred peak. The inferred peaks, although not strongly constrained, tend towards large scales ($0.5 R_{500}$ to $R_{500}$).
        
        \input{A3D_Mach_LinMix_CorrLogs}
        
        Table~\ref{tbl:A3D_M_correlations} reports correlations between either $A_{\rho}$ or $\mathcal{M}_{\rho}$ and the dynamical parameters cited above. We include an additional parameter, $\Delta R$ we define as:
        \begin{equation}
            \Delta R = \Delta_c / \theta_{500},
        \end{equation}
        where $\Delta_c$ is the angular distance between the SZ and X-ray centroids, and $\theta_{500}$ is the angular equivalent of $R_{500}$. We quantify the correlations with the Spearman and Pearson coefficients, $r_{\text{Sp}}$, and $r_{\text{Pe}}$, respectively. Additionally, we quantify the correlations via the linear correlation coefficient when considering a Bayesian approach with \lstinline{LINMIX}\footnote{as implemented in Python; see \url{https://linmix.readthedocs.io/}.} \citep{kelly2007,Gaspari2019}. This method takes observables $y$ and $x$, the relations $y = \eta + \sigma_y$, $x = \xi_c + \sigma_x$, where $\xi_c$ is the independent variable and $\eta$ is the dependent variable, and fits the linear relation:
        \begin{equation}
            \eta = \alpha_c + \beta_c \xi_c + \epsilon,
            \label{eqn:linmix}
        \end{equation}
        where $\epsilon$ is the intrinsic scatter and $\alpha_c$ and $\beta_c$ are the regression coefficients. The correlation coefficients from \lstinline{LINMIX} is that between $\xi_c$ and $\eta$ and is reported in Table~\ref{tbl:A3D_M_correlations} as $r_{\text{Lin}}$. We report the coefficients $\alpha_c$ and $\beta_c$ in Table~\ref{tbl:A3D_M_coefficients}.
        
        \input{A3D_Mach_Coefficients}

        \citet{dupourque2023} investigated correlations between the amplitude of fluctuations (related to the integral of the power spectrum of fluctuations), $\sigma_{\delta}$, and dynamical parameters $c$, $\omega$, the Gini coefficient, $G$, and an asymmetry parameter quantified through Zernike polynomials, $C_{Z}$.
        Although our comparisons are not precisely equivalent, we should expect that the correlations we find for $A_{\rho}(k_{\text{peak}}) - \log_{10}(c)$ are similar to those found in \citet{dupourque2023} for $\sigma_{\delta} - c$ and likewise for $A_{\rho}(k_{\text{peak}}) - \log_{10}(\omega)$ and their $\sigma_{\delta} - \omega$. This is in fact the case, where \citet{dupourque2023} find the Spearman coefficients for $\sigma_{\delta} - c$ and $\sigma_{\delta} - \omega$ to be $-0.4_{-0.15}^{+0.15}$ and $0.37_{-0.18}^{+0.2}$, respectively. Those values are similar to the analogous Spearman coefficients $-0.44^{+0.10}_{-0.08}$ and $0.48^{+0.08}_{-0.08}$ reported in Table~\ref{tbl:A3D_M_correlations}. In \citet{dupourque2024}, clusters were subdivided into three bins of dynamical state based on $\omega$ and a positive correlation with $\sigma_{\delta}$ is found, but no explicit calculation (correlation coefficient) is provided.

    \subsection{Correlations with the Mach number}

    Where Table~\ref{tbl:A3D_M_correlations} presented the correlations of dynamical parameters relative to both $A_{\rho}$ and $\mathcal{M}_{\rho}$ in Ring 1 
    %(noted here as $A_{\rho,1}$ and $\mathcal{M}_{\rho,1}$; i.e. indexing the Ring) 
    and Figure~\ref{fig:correlation1} visually presented correlations against $A_{\rho,1}$, Figure~\ref{fig:correlation2} does so for $\mathcal{M}_{\rho,1}$. As evidenced in Table~\ref{tbl:A3D_M_correlations}, the correlations do not differ drastically between the $A_{\rho,1}$ and $\mathcal{M}_{\rho,1}$ cases. %{\bf Additionally, of the dynamical parameters from \citet{yuan2022}, $\delta$ appears to have the best correlation with fluctuations, but the difference in correlation coefficients of other parameters is not so dramatic so as to}

    We considered additional correlations with $A_{\rho,1}$ or $\mathcal{M}_{\rho,1}$. There are several readily available quantities from previous works such as those in \citet{bulbul2019}, taken as the values within the aperture of $R_{500}$: $L_{\text{X,cin}}$, $L_{\text{X,cin,bol}}$, $T_{\text{X,cin}}$, $Z_{\text{X,cin}}$, $L_{\text{X,cex}}$, $L_{\text{X,cex,bol}}$, $T_{\text{X,cex}}$, $Z_{\text{X,cex}}$, $Y_{\text{X,cin}}$, $M_{500}$, and $z$, where subscripts cin and cex indicate whether the core ($r < 0.15 R_{500}$) is included or excised, respectively, for the quantities. We do not find any strong correlation amongst these variables and the inferred density fluctuations in Ring 1.  The lack of correlation between luminosities, temperatures, and integrated $Y$ relative to fluctuations is likely a symptom of the former quantities scaling with mass, whereas turbulence appears to be independent of mass \citep[e.g.][]{nelson2014a}, or have a mild dependence on mass \citep[e.g.][]{battaglia2012a,angelinelli2020}. Even considering ratios of core-included to core-excluded quantities does not produce any strong correlations; this reinforces the notion that such ratios are not robust tracers of dynamical state; e.g. cool cores can be present in both relaxed and disturbed systems.

    From the SPT-SZ works \citep{bleem2015,bocquet2019}, one also has $Y_{\text{SZ}}$. We therefore additionally consider the correlation of $Y_{\text{SZ}}/Y_{\text{X,cin}}$ and $A_{\rho,1}$, which yields no apparent correlation. Finally, we consider the self-similar scaling $Y_{\text{SZ}} \propto E(s)^{2/3} M^{5/3}$ \citep[e.g.][]{kravtsov2012} and compute a quantity $\Psi = Y_{\text{SZ}} / (E(s)^{2/3} M_{500}^{5/3})$ that we then correlate against $A_{\rho,1}$. This too does not show a clear correlation. Of quantities that we have correlated with fluctuations ($A_{\rho,1}$), or the inferred turbulent Mach numbers, $\mathcal{M}_{\rho,1}$, the quantities which we found to have statistically significant correlations are the dynamical parameters in \citet{yuan2022}. We do not find significant difference among the correlation coefficients of these dynamical parameters (for a given comparison, e.g. against $A_{\rho,1}$). More sensitive measurements, as well as a larger sample will help provide such distinctions.
    %Of course, $\Psi$ would be better calculated if $M_{500}$ were not itself derived from $Y_{SZ}$, e.g. if $M_{500}$ were derived from weak lensing.  

    \subsection{Distribution of fluctuations and inferred turbulent velocities}
    \label{sec:turb_dist}

        \begin{figure}
            \centering
            \includegraphics[width=0.47\textwidth]{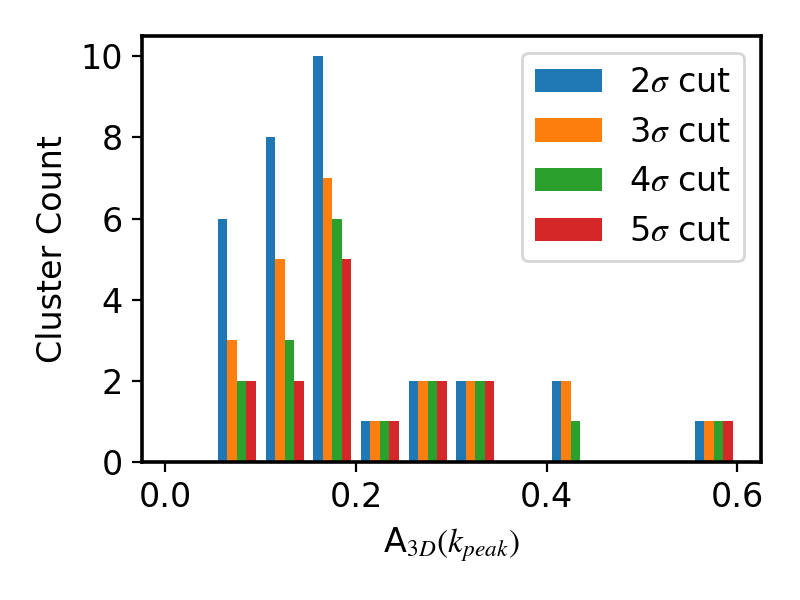}
            \includegraphics[width=0.47\textwidth]{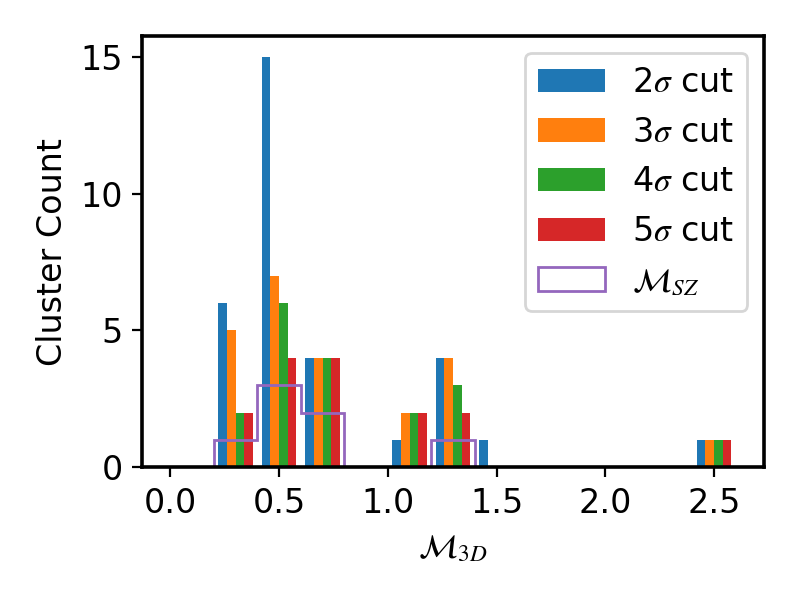}
            \caption{Within Ring 1, distributions of $A_{\rho}(k_{\text{peak}})$ and distributions of Mach numbers, $\mathcal{M}_{\rho}$ for different significance thresholds on $A_{\rho}$, and $\mathcal{M}_{P}$ with $A_{P}$ significance greater than $1.5\sigma$. For clarity, the blue bars indicate the number of clusters for which we infer a given peak of $A_{\rho}$ or a given $\mathcal{M}_{\rho}$ when considering only nodes of amplitude spectra for which $A_{\rho} > 2\sigma_{A_{\rho}}$.}
            \label{fig:distributions}
        \end{figure}

        The adopted threshold of $2\sigma$ ($\xi > 2$) is admittedly a low threshold and may introduce a bias due to noise that happens to scatter values above our threshold. Accounting for any bias is potentially quite involved, as there are at least two parts to consider: (1) what is the bias on the value of the amplitude spectrum at the considered wavenumber ($k_{\text{peak}}$), and (2) would a correction to this bias change the inferred $k_{\text{peak}}$? An earnest attempt to correct for this bias would require knowledge about the expected distribution of amplitude spectra, which is not yet established.

        To gauge the potential importance of such a bias, we investigate the inferred $A_{\rho}(k_{\text{peak}})$ using 2, 3, 4, and 5$\sigma$ cuts. We don't find (see Figure~\ref{fig:distributions}) clear evidence of a substantial bias in the distribution of $A_{\rho}(k_{\text{peak}})$. The respective weighted means of $A_{\rho}(k_{\text{peak}})$ with their statistical uncertainties (ignoring scatter) are $ 0.162 \pm 0.005, 0.156 \pm 0.004, 0.159 \pm 0.005, ~\text{and}~0.159 \pm 0.005$, thus revealing no significant tension.
        %Moreover, the weighted mean of $A_{\rho}(k_{\text{peak}})$ remains 0.16 across all significance cuts.
        
        \begin{figure}
            \centering
            \includegraphics[width=0.47\textwidth]{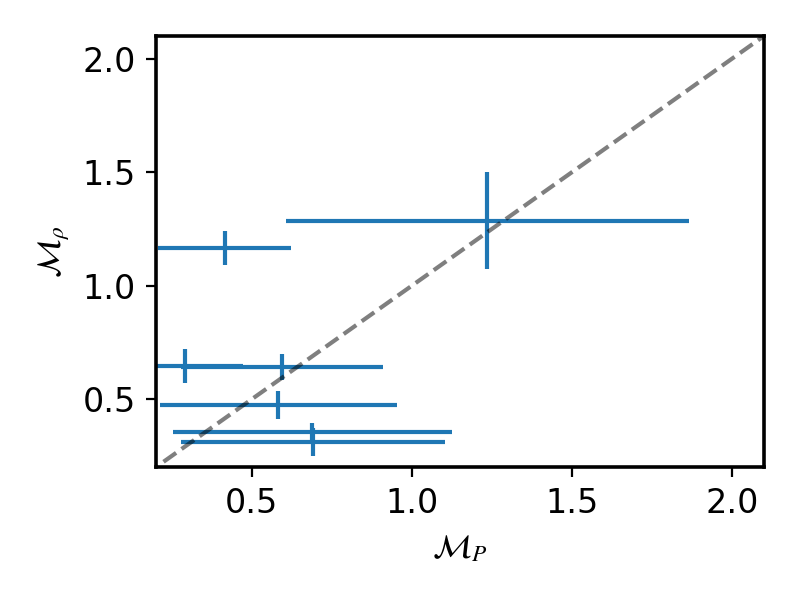}
            \caption{Comparison of $\mathcal{M}_{\text{3D}}$ as derived from SZ (pressure) or X-ray (density) fluctuations. The dashed line shows unitary equivalence.}
            \label{fig:Mach_comparison}
        \end{figure}

        While $A_{\rho}(k_{\text{peak}})$ appears to not suffer a substantial bias, we may also be concerned with the inferred turbulent velocity. Again, we do not see evidence for a clear bias from the $2\sigma$ cut (see again Figure~\ref{fig:distributions}), where the weighted means are 0.57, 0.57, 0.59, and 0.59 for the respective threshold cuts of 2, 3, 4, and 5$\sigma$. Across the thresholds, we can thus find an average turbulent velocity $\mathcal{M}_{\rho} \sim 0.6$. Additionally, the bottom panel of Figure~\ref{fig:distributions} is suggestive of an underlying bimodal distribution, where the two populations are separated at the supersonic transition. Using the dip test from \citet{hartigan1985} on our distribution of Mach numbers, we find dip values less than 0.07 across the threshold cuts, which correspond to probabilities, $p$, of a unimodal distribution $0.48 < p < 0.84$. For the distribution of $A_{\text{3D}}$ peaks, (across the cuts) we find $p > 0.95$. If the distributions are indeed unimodal, they are positively skewed, where the moment of skewness (across all significance cuts) for $A_{\text{3D}}$ peaks is $\geq 1.3$, while for $\mathcal{M}_{\text{3D}}$ the values are $\geq 1.6$.

        To add to these tests, we also investigate a Gaussian mixture and use the changes in the Bayesian information criterion (BIC) between the case of two Gaussians ($\text{BIC}_{2}$) and the case of one Gaussian ($\text{BIC}_{1}$), such that $\Delta \text{BIC} = \text{BIC}_{2} - \text{BIC}_{1}$. For the distribution of $A_{\text{3D}}$ peaks, we find $\Delta \text{BIC} = -14.0,-5.0,-0.9,$ and $1.6$ for the respective threshold cuts of 2, 3, 4, and 5$\sigma$. The analogous values for the $\mathcal{M}_{\text{3D}}$ distributions are $\Delta \text{BIC} = -18.2,-6.9,-4.4,$ and $-10.9$. \citet{Kass1995} indicates that $\Delta \text{BIC} < -6$ provides strong evidence that, in this case, the underlying distribution is best described by two Gaussians rather than a single Gaussian. Thus, while the diptest suggests a unimodal distribution (as in there is not a robust trough in the distribution), we can confirm that any such unimodal distribution is not well described by a single Gaussian.

        %https://pypi.org/project/diptest/

         We also consider the weighted means if we trim the clusters with supersonic Mach numbers. Doing so, we find weighted means (of the Mach numbers) of: 0.52, 0.50, 0.52, and 0.52 for the respective $\sigma$ cuts. The respective statistical uncertainties of these values are 0.019, 0.016, 0.017, and 0.018, indicating no substantial bias. Such statistical uncertainties ignore intrinsic scatter, which is between 0.12 and 0.14 in all four cases. The median subsonic $\mathcal{M}_{\rho}$ values for the respective $\sigma$ cuts are 0.52, 0.50, 0.55, and 0.56. In the case of the $2\sigma$ cut, there are 25 clusters with inferred turbulent velocities that are subsonic.  When restricting velocities to those which are subsonic, across the thresholds, we find an average $\mathcal{M}_{\rho} \sim 0.5$
         In the following section we discuss the interpretation of supersonic Mach numbers and identify the individual clusters in which we infer supersonic velocities.\\ %such numbers.
         %In the following section we note properties of individual clusters which are identified as having supersonic turbulent Mach numbers.

         For the seven clusters which have SZ constraints (above $1.5\sigma$ in Ring 1), we compare the SZ-inferred turbulent velocities to those from X-ray, i.e. we compare the inferences from pressure and density fluctuations, in Figure~\ref{fig:Mach_comparison}. We find general agreement and note the point with $\mathcal{M}_{\rho} > 1$ and $\mathcal{M}_{\text{P}} < 0.5$ corresponds to SPT-CLJ0014-3022 (or Abell 2744). Notwithstanding considerations of masking substructure (see Apendix~\ref{sec:substructure_masking}), this indicates that the infalling group has not contributed to substantial pressure fluctuations.

        %We also include the distribution of Mach numbers as inferred from SZ data (i.e.~pressure fluctuations) in Figure~\ref{fig:distributions}. The Mach number for the Bullet cluster is recovered as $2.3 \pm 1.2$, and is not included in this distribution. 

%% file: A3D_Mach_LinMix_CorrLogs.tex
\begin{deluxetable*}{ccccccc} 
\tablecaption{Correlation coefficients \label{tbl:A3D_M_correlations}}   
%\tabletypesize{\footnotesize}  % \small, \footnotesize, or \scriptsize
\tablehead{
\colhead{Dynamical} & \multicolumn{3}{c}{$\log{ A_{\rho} }$}  & \multicolumn{3}{c}{$\log{ \mathcal{M}_{\rho} }$} \\[-0.25cm]
\colhead{Parameter} & \colhead{$r_{\text{Sp}}$} & \colhead{$r_{\text{Pe}}$} & \colhead{$r_{\text{Lin}}$} & \colhead{$r_{\text{Sp}}$} & \colhead{$r_{\text{Pe}}$} & \colhead{$r_{\text{Lin}}$}
} 
\startdata 
%$\log{c}$ & $-0.44^{+0.09}_{-0.10}$ & $-0.49^{+0.08}_{-0.07}$ & $-0.53^{+0.17}_{-0.16}$ & $-0.39^{+0.05}_{-0.04}$ & $-0.53^{+0.02}_{-0.02}$ & $-0.45^{+0.17}_{-0.15}$ \\ 
%$\log{\omega}$ & $0.48^{+0.07}_{-0.08}$ & $0.49^{+0.05}_{-0.06}$  & $0.55^{+0.15}_{-0.17}$ & $0.45^{+0.04}_{-0.04}$ & $0.53^{+0.01}_{-0.02}$ & $0.44^{+0.15}_{-0.17}$ \\ 
%$\log(\frac{P_3}{P_0})$ & $0.35^{+0.11}_{-0.11}$ & $0.36^{+0.08}_{-0.09}$ & $0.47^{+0.17}_{-0.20}$ & $0.39^{+0.07}_{-0.08}$ & $0.41^{+0.05}_{-0.05}$ & $0.44^{+0.16}_{-0.18}$ \\ 
%$\kappa$ & $0.44^{+0.08}_{-0.10}$ & $0.44^{+0.06}_{-0.08}$ & $0.46^{+0.16}_{-0.18}$ & $0.30^{+0.04}_{-0.04}$ & $0.30^{+0.02}_{-0.02}$ & $0.32^{+0.16}_{-0.18}$ \\ 
%$\log{\alpha}$ & $0.45^{+0.08}_{-0.10}$ & $0.44^{+0.07}_{-0.07}$ & $0.52^{+0.17}_{-0.18}$  & $0.55^{+0.04}_{-0.04}$ & $0.54^{+0.01}_{-0.01}$ & $0.45^{+0.15}_{-0.17}$ \\ 
%$\delta$ & $0.54^{+0.08}_{-0.09}$ & $0.51^{+0.05}_{-0.06}$ & $0.54^{+0.15}_{-0.17}$  & $0.55^{+0.04}_{-0.04}$ & $0.54^{+0.01}_{-0.01}$ & $0.44^{+0.15}_{-0.17}$ \\ 
%$\Delta R$ & $-0.06^{+0.10}_{-0.10}$ & $-0.06^{+0.07}_{-0.07}$ & $-0.12^{+0.21}_{-0.21}$  & $-0.14^{+0.05}_{-0.05}$ & $-0.11^{+0.03}_{-0.03}$ & $-0.10^{+0.19}_{-0.19}$ 
%
% Now A, and B are the last two columns. This was the Mach correlations
$\log{c}$ & $-0.44^{+0.10}_{-0.08}$ & $-0.43^{+0.09}_{-0.07}$ & $-0.60^{+0.17}_{-0.14}$ & $-0.34^{+0.05}_{-0.05}$ & $-0.37^{+0.03}_{-0.04}$ & $-0.39^{+0.19}_{-0.16}$ \\ 
$\log{\omega}$ & $0.48^{+0.08}_{-0.08}$ & $0.44^{+0.06}_{-0.07}$ & $0.59^{+0.13}_{-0.16}$ & $0.40^{+0.04}_{-0.04}$ & $0.40^{+0.02}_{-0.02}$ & $0.42^{+0.15}_{-0.17}$ \\ 
$\log(\frac{P_3}{P_0})$ & $0.34^{+0.11}_{-0.12}$ & $0.34^{+0.10}_{-0.12}$ & $0.50^{+0.17}_{-0.20}$ & $0.34^{+0.08}_{-0.09}$ & $0.35^{+0.07}_{-0.08}$ & $0.40^{+0.17}_{-0.19}$  \\ 
$\kappa$ & $0.44^{+0.08}_{-0.10}$ & $0.41^{+0.08}_{-0.09}$ & $0.56^{+0.15}_{-0.17}$ & $0.27^{+0.04}_{-0.04}$ & $0.30^{+0.03}_{-0.03}$ & $0.31^{+0.18}_{-0.19}$ \\ 
$\log{\alpha}$ & $0.43^{+0.09}_{-0.09}$ & $0.42^{+0.08}_{-0.09}$ & $0.57^{+0.15}_{-0.19}$ & $0.51^{+0.04}_{-0.05}$ & $0.44^{+0.03}_{-0.03}$ & $0.45^{+0.15}_{-0.17}$ \\ 
$\delta$ & $0.54^{+0.08}_{-0.09}$ & $0.49^{+0.08}_{-0.09}$ & $0.67^{+0.13}_{-0.17}$ & $0.44^{+0.04}_{-0.04}$ & $0.42^{+0.03}_{-0.03}$ & $0.44^{+0.15}_{-0.18}$  \\ 
$\Delta R$ & $-0.06^{+0.10}_{-0.10}$ & $-0.03^{+0.09}_{-0.09}$ & $-0.09^{+0.22}_{-0.22}$ & $-0.14^{+0.06}_{-0.05}$ & $-0.08^{+0.04}_{-0.04}$ & $-0.08^{+0.20}_{-0.19}$  
\enddata 
\tablecomments{Correlation coefficients obtained between either $\log{A_{\rho}(k_{\text{peak}})}$ or $\log{A_{\rho}(k_{\text{peak}})}$ and various dynamical parameters with the cut $\xi_{A_{\rho}} > 2$.} 
\end{deluxetable*} 

%% file: A3D_Mach_Coefficients.tex
\begin{deluxetable}{ccccc} 
\tablecaption{Linear coefficients \label{tbl:A3D_M_coefficients}}   
\tabletypesize{\scriptsize}  % \small, \footnotesize, or \scriptsize
\tablehead{
\colhead{Dynamical} & \multicolumn{2}{c}{$\log{ A_{\rho} }$}  & \multicolumn{2}{c}{$\log{ \mathcal{M}_{\rho} }$} \\[-0.25cm]
\colhead{Parameter} & \colhead{$\alpha_c$} & \colhead{$\beta_c$} &  \colhead{$\alpha_c$} & \colhead{$\beta_c$} 
} 
\startdata 
$\log{c}$ &  $-1.23^{+0.17}_{-0.16}$ & $-0.69^{+0.22}_{-0.21}$ & $-0.79^{+0.06}_{-0.05}$ & $-0.35^{+0.17}_{-0.17}$ \\ 
$\log{\omega}$ & $-0.47^{+0.47}_{-0.47}$ & $1.97^{+0.62}_{-0.62}$ & $-1.72^{+0.16}_{-0.15}$ & $1.07^{+0.46}_{-0.45}$ \\ 
$\log(\frac{P_3}{P_0})$ & $-5.11^{+0.60}_{-0.59}$ & $1.87^{+0.80}_{-0.76}$ & $-6.25^{+0.18}_{-0.20}$ & $1.25^{+0.61}_{-0.62}$ \\ 
$\kappa$ & $2.74^{+0.36}_{-0.36}$ & $1.44^{+0.48}_{-0.47}$ & $1.81^{+0.13}_{-0.13}$ & $0.68^{+0.41}_{-0.41}$ \\ 
$\log{\alpha}$ & $-0.26^{+0.33}_{-0.32}$ & $1.22^{+0.42}_{-0.41}$ & $-1.01^{+0.10}_{-0.10}$ & $0.78^{+0.31}_{-0.31}$ \\ 
$\delta$ & $2.04^{+0.39}_{-0.39}$ & $1.88^{+0.51}_{-0.51}$ & $0.84^{+0.14}_{-0.13}$ & $1.03^{+0.42}_{-0.41}$ \\ 
$\Delta R$ & $0.06^{+0.06}_{-0.06}$ & $-0.03^{+0.08}_{-0.09}$ & $0.09^{+0.02}_{-0.02}$ & $-0.03^{+0.06}_{-0.06}$ 
\enddata 
\tablecomments{Linear coefficients $\alpha_c$ and $\beta_c$, given in Equation~\ref{eqn:linmix} obtained between $\log{A_{\rho}(k_{\text{peak}})}$ and various dynamical parameters with the cut $\xi_{A_{\rho}} > 2$.} 
\end{deluxetable} 

%% file: Discussion.tex
    %%% I think this already says it, but I can explicitly use the terminology of "turbulence-driven" or "turbulence-dominated" versus "shock-dominated"

    In the previous section we found an average turbulent velocity within Ring 1 ($R < 0.62 R_{500}$) which corresponds to $\mathcal{M}_{\rho} \sim 0.6$ when including all clusters across the considered significance cuts in the amplitude spectrum of density fluctuations. This average becomes $\mathcal{M}_{\rho} \sim 0.5$ when confining attention to the subset of clusters that also have inferred $\mathcal{M}_{\rho} < 1$.%(from Section~\ref{sec:turb_dist}: $\mathcal{M}_{\rho} = 0.52 \pm 0.14$ when taking $\xi > 2$ and clusters wherein $\mathcal{M}_{\rho} < 1$). 
    
    Given that turbulence with $\mathcal{M}_{\text{3D}} > 1$ is largely not expected within $R_{500}$, let alone within $0.62 R_{500}$ and that the clusters for which we infer $\mathcal{M}_{\rho} > 1$ have either known merger shocks or morphologies suggestive of mergers (see Section~\ref{sec:Supersonic}) we consider that the inferred Mach numbers should not be interpreted as arising solely from turbulence. We discuss this more in the following subsection.

    Previous studies of surface brightness fluctuations across samples of similar sizes have tended to find relatively lower 3D Mach numbers than the $\mathcal{M}_{\rho} = 0.52 \pm 0.14$ found in this work (using clusters with subsonic inferred turbulent velocities and $\xi > 2$; see Section~\ref{sec:turb_dist}). 
    For example, \citet{hofmann2016} find an average $\mathcal{M}_{\rho} \approx 0.3$ with a large 50\% scatter in a sample of 33 \textit{Chandra} clusters.
    Across the 12 clusters in the small X-COP sample, \citet[][]{dupourque2023} found $\mathcal{M}_{\rho} = 0.37 \pm 0.06$ within $0.5 R_{500} < r < R_{500}$.
%    $\mathcal{M}_{\rho} = 0.1 \pm 0.01$, $0.15 \pm 0.02$, $0.1 \pm 0.02$, and $0.37 \pm 0.06$ in the regions $r < 0.1 R_{500}$, $0.1 R_{500} < r < 0.25 R_{500}$, $0.25 R_{500} < r < 0.5 R_{500}$, and, $0.5 R_{500} < r < R_{500}$, respectively. 
    Investigating a sample of 80 clusters with \textit{Chandra} data, \citet[][]{heinrich2024}
    found $\mathcal{M}_{\rho} = 0.31 \pm 0.09$ in the region within $0.4 R_{500}$.  Using 64 (of the 82) clusters in the CHEX-MATE sample, \citet[][]{dupourque2024} find an average turbulent velocity of $\mathcal{M}_{\rho} = 0.41 \pm 0.17$ within $R_{500}$. On 28 of the CHEX-MATE clusters, \citet{Lovisari2024} find 17\% temperature fluctuations and infer $\mathcal{M}_{\text{3D}} = 0.37_{-0.09}^{+0.16}$.

    While sample selection may play into differences in inferred turbulent velocities, neither our sample nor those in other works can be strictly described as SZ- or X-ray selected samples. Our treatments of masking substructure or excluding merging systems are also heterogeneous. We investigate differences in sample distributions in Appendix~\ref{sec:app_table}
    %The larger difference  in average Mach number with pure X-ray samples (e.g.~\textit{Chandra} samples) can be understood as the SZ selection is typically less biased, e.g., allowing us to avoid the cool-core bias in X-ray band \citep{Rossetti2017}. This enables the inclusion of a larger fraction of substantially perturbed and unrelaxed systems (with boosted pressure/SZ signal within $R_{2500}-R_{500}$), leading to a net increase in the observed average Mach number.

    To better understand/test the Mach numbers, we compare them to predictions from cosmological hydrodynamical simulations. \citet{battaglia2012a} (B12), \citet{nelson2014a} (N14), and \citet{angelinelli2020} (A20) have investigated non-thermal pressure profiles, $P_{\text{NT}}$, due to random or kinetic motions, where \citet{angelinelli2020} provides an explicit separation for the pressure due to strictly turbulent motions. In particular, these works provide parametric forms for $\alpha_{\text{NT}} = P_{\text{NT}} / (P_{\text{NT}} + P_{\text{th}})$, where $P_{\text{th}}$ is the thermal pressure. For turbulent motions, $P_{\text{turb}}/P_{\text{th}} = (\gamma/3)\mathcal{M}_{\text{3D}}^2$. Thus, where $P_{\text{NT}}$ is taken to be, implicitly or explicitly, $P_{\text{turb}}$, one can infer $\mathcal{M}_{\text{3D}}$. 
    %Although these simulations were not designed to model (or resolve) intricacies deep into the cluster cores, but are expected to capture gas motions at moderate cluster radii reasonably well. 
    
    \begin{figure}
        \centering
        \includegraphics[width=0.47\textwidth]{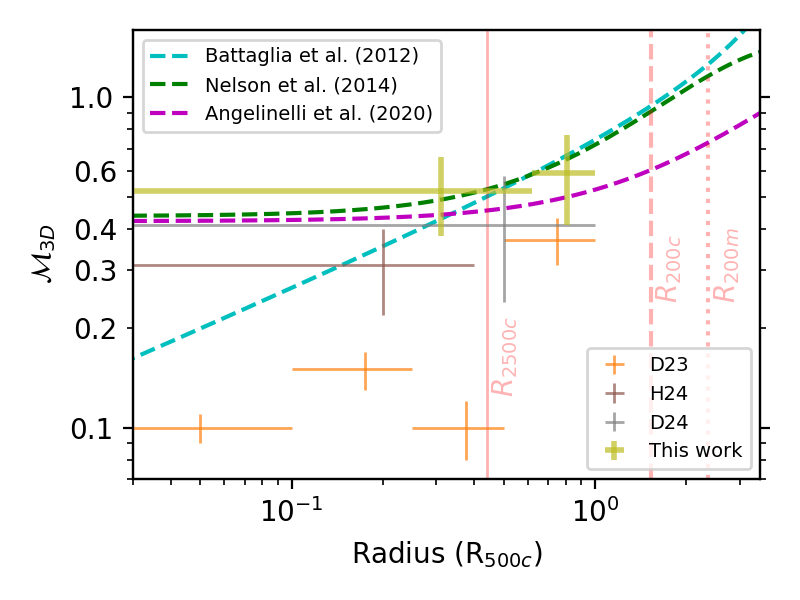}
        \caption{Radial profiles of Mach numbers from various simulations (dashed curves) and observational constraints on Mach numbers as points with uncertainties (in $\mathcal{M}_{\text{3D}}$); radial "error bar" denotes the extent of the radial bin. Observational points identified as D23, H24, and D24 refer to \citet{dupourque2023}, \citet{heinrich2024}, and \citet{dupourque2024}, respectively. Points from this work correspond to those in which gas motions have been restricted to subsonic velocities.}
        \label{fig:Mach_profiles}
    \end{figure}
    
    Figure ~\ref{fig:Mach_profiles} shows Mach profiles derived from the non-thermal pressure profiles presented in B12, N14, and A20, where we take the profile explicitly determined for turbulence from A20.
    From these profiles we further calculate that within $0.62 R_{500}$, the expected $\mathcal{M}_{\text{3D}}$ values are 0.49, 0.52, and 0.45 for B12, N14, and A20, respectively. Within Ring 2 ($0.62 R_{500} < r < R_{500}$) those respective Mach numbers are 0.68, 0.66, and 0.50. We note that the simulations themselves find a scatter of $\gtrsim 10$\% in the $\alpha_{\text{NT}}$ profiles.

    Aside from differences in cluster samples and analysis approaches among \citet{dupourque2023,dupourque2024}, \citet{heinrich2024}, and this work, there are differences in the scaling between density fluctuations and inferred Mach numbers which are potentially relevant. While some theoretical agreement has been found between the relations presented in \citet{Gaspari2013_PS,Gaspari2014_PS} and \citet{zhuravleva2014,zhuravleva2023}, it will be important to establish a robust, empirical relation between density (and pressure) fluctuations and turbulent velocities inferred from high-resolution X-ray spectroscopy, e.g.~with the ongoing \textit{XRISM} mission.

    \subsection{Non-thermal pressure support and hydrostatic mass bias}
    \label{sec:NTP}

        Our inferred average turbulent Mach number, $\mathcal{M}_{\rho} = 0.52 \pm 0.14$ in Ring 1 is thus in excellent agreement with what is expected from simulation. This turbulent Mach number corresponds to a non-thermal pressure fraction $\alpha_{\text{NT}} = 0.13 \pm 0.06$. This value does not necessarily reflect the hydrostatic mass bias, which is instead given by:
        \begin{equation}
            b_{\mathcal{M}} = \frac{-\gamma \mathcal{M}_{\text{3D}}^2}{3} \frac{ d \ln P_{\text{NT}}}{d \ln P_{\text{th}}} \left( 1 +  \frac{\gamma \mathcal{M}_{\text{3D}}^2}{3}\frac{ d \ln P_{\text{NT}}}{d \ln P_{\text{th}}} \right)^{-1}
            \label{eqn:mach_bias}
        \end{equation}
        \citep[][]{khatri2016,romero2024}. That is, $\alpha_{\text{NT}} = -b_{\mathcal{M}}$ only when $d \ln P_{\text{NT}}/d \ln P_{\text{th}} = 1$. Considering that
        \begin{equation}
            \frac{ d \ln P_{\text{NT}}}{d \ln P_{\text{th}}} = 1 + 2 \frac{ d \ln \mathcal{M}_{\text{3D}} / d \ln r}{d \ln P_{\text{th}} / d \ln r},
        \end{equation}
        we see that $d \ln P_{\text{NT}}/d \ln P_{\text{th}} = 1$ when the turbulent Mach number is constant with radius. We expect that $\mathcal{M}_{\text{3D}}$ should, in general, increase with radius (as in Figure~\ref{fig:Mach_profiles}) and consequently we expect that $-b_{\mathcal{M}} < \alpha_{\text{NT}}$.  
    
    %We should specify that this relation should be understood in the context of an effective radius at which these values are calculated. In our case, $\alpha_{\text{NT}} = 0.13 \pm 0.06$ applies to the region $r < 0.62 R_{500}$. 
    %where we may take the effective radius of Ring 1 to be $R_{\text{eff},1} = 0.44 R_{500}$. 

        From above, we can say that we expect the average hydrostatic mass bias within Ring 1 to be less than $0.13$ (for those clusters with inferred subsonic turbulence). Given that masses are generally not provided at $0.62 R_{500}$, the hydrostatic mass bias at $R_{500}$ is of more interest. However, our constraints from Ring 2 are poorer and we discuss these in detail in Appendix~\ref{sec:app_ring2}. 
        %Here, we briefly pull relevant results from Ring 2. 
        From the few (four) clusters which have sufficient constraints (at least $2\sigma$ significance in $A_{\rho,2}$) and yielding subsonic turbulent velocities, we find $\mathcal{M}_{\rho,2} = 0.59 \pm 0.18$ with corresponding $\alpha_{\text{NT}} = 0.16 \pm 0.08$. We can take this to define an upper limit of the hydrostatic mass bias: $-b_{\mathcal{M}} < 0.16 \pm 0.08$. Such an interpretation is consistent with the expected hydrostatic mass bias values between 0.1 and 0.3 \citep[e.g.][and references therein]{romero2024}; though we note again that the sample size is small (four clusters) and the constraints are of limited quality. This motivates our later discussion in Section\,\ref{sec:toward}.
        %If we include those with supersonic velocities (totaling fifteen clusters), we find $\mathcal{M}_{\rho,2} = 0.87 \pm 0.89$ with a corresponding  $\alpha_{\text{NT}} = 0.59 \pm 0.26$. 

    %Unfortunately, an estimate of $-b_{\mathcal{M}}(R_{500})$

    \subsection{Inferred supersonic velocities}
    \label{sec:Supersonic}

        The clusters for which $\mathcal{M}_{\rho} > 1$ are: SPT-CLJ0354-5904, SPT-CLJ0658-5556, SPT-CLJ2017-6258, SPT-CLJ2056-5459, SPT-CLJ0304-4401, SPT-CLJ2032-5627, and SPT-CLJ0014-3022. Of these, SPT-CLJ0354-5904, SPT-CLJ2017-6258, and SPT-CLJ2056-5459 show asymmetries or potential substructure in the \textit{XMM} images that is suggestive of disturbance. Dynamical parameters from \cite{yuan2022} for these three clusters corroborate this. For example, all three of these clusters have profile parameters, $\kappa > 2$, and of the 32 clusters with sufficient constraints, these three are among the top five clusters with respect to highest values of $\kappa$. However, our substructure algorithm did not identify any substructure to mask, in part due to the modest photon counts in those images. SPT-CLJ0658-5556 (the Bullet cluster), SPT-CLJ0014-3022 (Abell 2744), SPT-CLJ0304-4401, and SPT-CLJ2032-5627 are known mergers \citep{markevitch2002,kempner2004,raja2021,duchesne2021}. 

        In the known mergers, we find that known or plausible shocks are within the same regions in which we infer $\mathcal{M}_{\rho} > 1$, consistent with our findings in \citet{romero2024}.
        %As in \citet{romero2024}, the regions where $\mathcal{M}_{\rho}$ is inferred, from surface brightness fluctuations, to exceed unity is broadly congruent with where known or plausible shocks exist. 
        Our analysis has assumed that density and pressure fluctuations, $A_{\text{3D}}$, scale linearly with the Mach number, which should hold for distributed turbulence (Section~\ref{sec:intro}). However, the inferred density and pressure fluctuations represent a volume-weighted average that can be accentuated due to super-linear levels by the influence of local shock(s), which are inherently supersonic. 
        We note that by super-linear fluctuations, we mean that $A_{\text{3D}}(k_{\text{peak}}) \propto \mathcal{M}_{\text{3D}}^x$ with $x > 1$. Such behavior due to shocks within a region could explain the skewed, if not bimodal, distribution of our inferred Mach numbers (in Figure~\ref{fig:distributions}).
        %Where we expect the linear relation ($A_{\text{3D}}(k_{\text{peak}}) \propto \mathcal{M}_{\text{3D}}$) for turbulence, this relation likely doesn't hold in the vicinity of shocks, where the Rankine-Hugoniot jump conditions would suggest $x \approx 2$. Thus, a behavior, intertwined with the distribution of shock strengths, could explain the skewed, if not bimodal, distribution of our inferred Mach numbers (in Figure~\ref{fig:distributions}).}
        From another perspective, someone could select a target cluster from super-linear fluctuations, and then investigate the (likely) presence of shocks with deeper observations.
        
        A more detailed interpretation of these supersonic velocities is likely to be complicated by several factors. As is often the case, the inclination angle of features, in this case shocks or a sloshing core, will impact the surface brightness signature. The current method of inferring gas velocities is developed in the context of turbulent motions and does not explicitly account for such substructure and thus different inclination angles. As such, we acknowledge that our inferred volume-averaged gas velocities have additional (unaccounted) systematic uncertainties. Secondarily, there is the matter of masking, which has evaded a widely accepted identification strategy \citep[e.g.][and this work]{Zhuravleva2015,dupourque2023}. The Bullet cluster and Abell 2744 provide some insight here, insofar as it is clear that our masking algorithm has masked the cooler gas behind the shocks in those two clusters, and \textit{not} the shocks themselves.

    \subsection{Towards more sensitive measurements} \label{sec:toward}

        This project aimed to constrain both pressure and density fluctuations, ideally out to $R_{500}$, over a sample of galaxy clusters with both sensitive X-ray and SZ data. We find that it is already difficult to place tight constraints on these fluctuations within $0.62 R_{500}$. On the SZ side, pressure fluctuation constraints are at best $2\sigma$. The ongoing SPT-3G survey \citep{benson2014} is expected to reach a final depth 10 times that of the SPT-SZ survey, and correspondingly, we should expect the uncertainties in amplitude spectra to improve by a factor of 10 except for the nodes at largest scales, which may become dominated by cosmic variance \citep[e.g.][]{romero2024b}. This will enable some insight into pressure fluctuations, but the constraints across spatial scales will still be limited, due to the expected power spectrum of pressure fluctuations and the angular resolution achieved by SPT-3G.

        With respect to the dependence of the SNR on the angular frequency, $k$, we can take a simple case where the statistical noise in $y$ or $S$ maps has a flat power spectrum. In this case, the uncertainty in the measurements of surface brightness fluctuations will scale as $k^{-1}$ \citep{arevalo2012,romero2024b}. At scales smaller than the injection scale, the power spectrum will have a logarithmic slope steeper than $-3$, where at some frequency beyond $k_{\text{inj}}$, a logarithmic slope of $11/3$ is predicted for Kolmogorov turbulence. Therefore, we can consider that the SNR is declining at a rate steeper than $k^{-2}$ beyond the injection scale.

        While the resolution of \textit{XMM-Newton} has a non-trivial impact on measurements of density fluctuations, we see that, in fact, constraints at the smallest scales tend to still be limited by overall sensitivity owing to the scaling of SNR with $k$. Consider that we do achieve $> 3\sigma$ constraints at spatial scales corresponding to $\sim 0.4 R_{500}$, which appears to be a plausible injection scale (e.g.~\citealt{Gaspari2014_PS}). Suppose a cluster has a constraint of $3\sigma$ at $0.4 R_{500}$ and we wish to obtain $3 \sigma$ constraints at $0.1 R_{500}$; we would need to improve the SNR by at least 16 times, which will require a factor of $16^2 = 256$ more time. 

        When we consider how the required time scales, this suggests clean exposure time requirements with \textit{XMM-Newton} in excess of 1 Ms (per cluster) to achieve $3\sigma$ constraints at $0.1 R_{500}$, in Ring 1 across our sample. From the constraints we do have in Ring 2, we find that the clean exposure time requirements exceed 10 Ms (per cluster; sometimes exceeding 100 Ms) for the same target constraint. It is clear that such constraints must be tasked to future facilities. 

        We would be remiss to not mention constraints to come from high-resolution spectroscopy, especially from \textit{XRISM} \citep{xrism2020} and the proposed \textit{Athena} \citep{barret2020,NewAthena2025NatAs} telescope. These will clearly play an important role in constraining turbulent motions in clusters. While there has not been a study comparing required observing times across spectroscopic instruments, nor a dedicated study comparing constraints from spectroscopy to those from surface brightness fluctuations, \citet{romero2024b} found that, for similar collecting areas, constraints from surface brightness fluctuations can be obtained with exposures that are one to two orders of magnitude shorter than those from spectroscopy\footnote{This comparison was made between the results in \citet{romero2024b} and those in \citet{beaumont2024}. As the differences in the methodologies are not trivial, a more judicious comparison between constraints from surface brightness fluctuations and high-resolution spectroscopy is warranted.}.
        
        %can be obtained with significantly shorter exposures.

%% file: Conclusions.tex
Expanding on our previous pilot investigations of surface brightness fluctuations jointly in the SZ and X-ray \citep{khatri2016,romero2023,romero2024}, we analyzed here, for the first time, a large sample of 60 clusters using both
SPT and \textit{XMM-Newton} data.
We thus derived constraints on density and pressure fluctuations within $\sim$\,$0.6 R_{500}$ for 32 and 7 clusters, respectively, and converted them to 3D Mach numbers through the peak amplitude linear relation (\citealt{Gaspari2013_PS}). 
We were able to derive constraints on density fluctuations out to $R_{500}$ for 15 clusters, but the interpretation of those fluctuations is unclear. We thus focus on the interpretation of fluctuations within $0.62\,R_{500}$;
our main results tied to this region are as follows.
\begin{itemize}
\item We find a mean Mach number to be $\mathcal{M}_{\rho} = 0.52 \pm 0.14$ for the 25 clusters that we consider to be dominated by turbulence and consistent with expectations from simulations \citep{battaglia2012a,nelson2014a,angelinelli2020}.
\item Clusters with supersonic $\mathcal{M}_{\rho} > 1$ are either known to be or are plausibly undergoing a merger, thus likely shock-dominated systems
\item For clusters with constraints on both density and pressure fluctuations, the inferred velocities are generally in agreement (except in Abell\,2744). %(where $\mathcal{M}_{\rho} > \mathcal{M}_{P}$).
\item We find mild correlations between the spectral amplitude/Mach number and the cluster dynamical parameters  that are typically used in the literature. 
\item Conversely, we find no significant correlation between the spectral amplitude/Mach number and either cluster mass or redshift, which is consistent with other observational and theoretical studies.
%For parameters also explored in \citet{dupourque2023}, we find consistency in our correlations.
\end{itemize}

In order to obtain robust results out to $R_{500}$, as well as tracing the full turbulent cascade, deeper observations are required. Some meaningful improvements are in progress, with SPT-3G, and can be obtained with deep \textit{XMM-Newton} observations. However, for the less massive and higher redshift clusters, robust constraints must come from future generation of instruments, both in X-ray and SZ band.

%% file: Appendix_Table.tex
%\begin{landscape}
    Tables~\ref{tbl:sample_good} and \ref{tbl:sample_bad} list various observational properties of the clusters. Clusters with density fluctuations in Ring 1, $A_{\rho,1}$, with at least one node of SNR $\xi > 2$ are listed in Table~\ref{tbl:sample_good} along with key properties of $A_{\rho,1}$. Conversely, clusters which do not satisfy the SNR threshold are listed in Table~\ref{tbl:sample_bad}. SPT-CLJ2344-4243 (Phoenix cluster), SPT-CLJ0637-4829, SPT-CLJ0330-5228, and SPT-CLJ2332-5358 are omitted from these tables. In the case of SPT-CLJ2344-4243, the bright cool-core and asymmetries of the \textit{XMM-Newton} PSF present a challenge beyond the scope of this work. The remaining three clusters suffered from complications in ESAS processing.

    \subsection{Cluster distribution}

\begin{figure}
            \centering
            \includegraphics[width=0.47\textwidth]{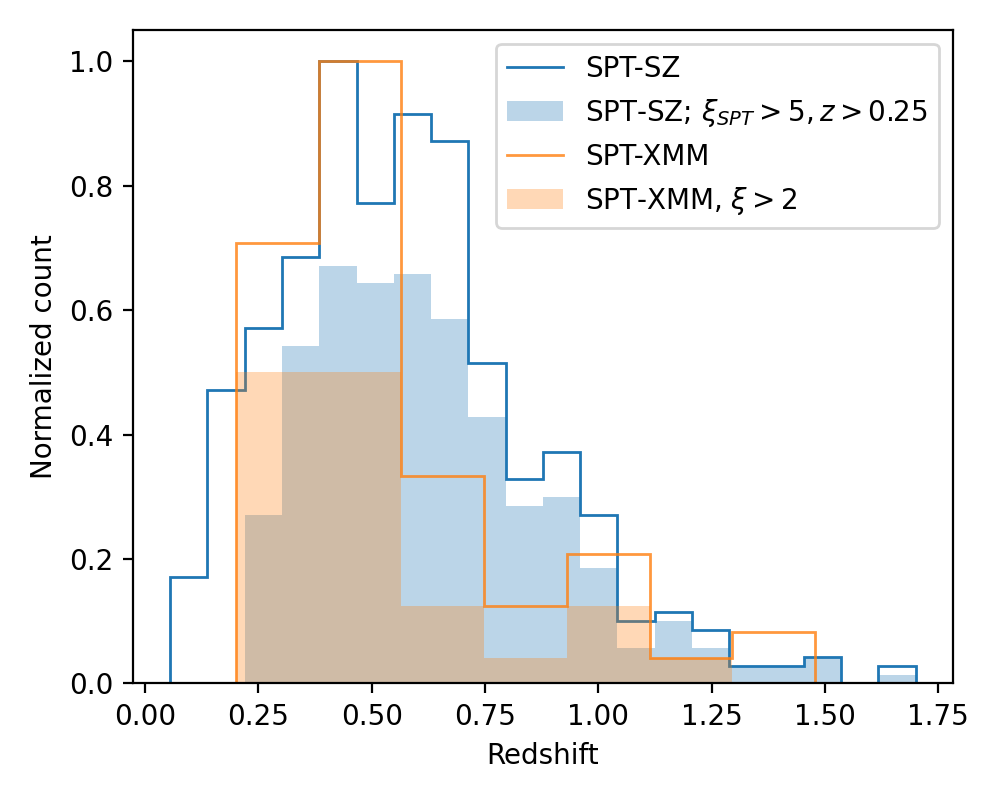}
            \includegraphics[width=0.47\textwidth]{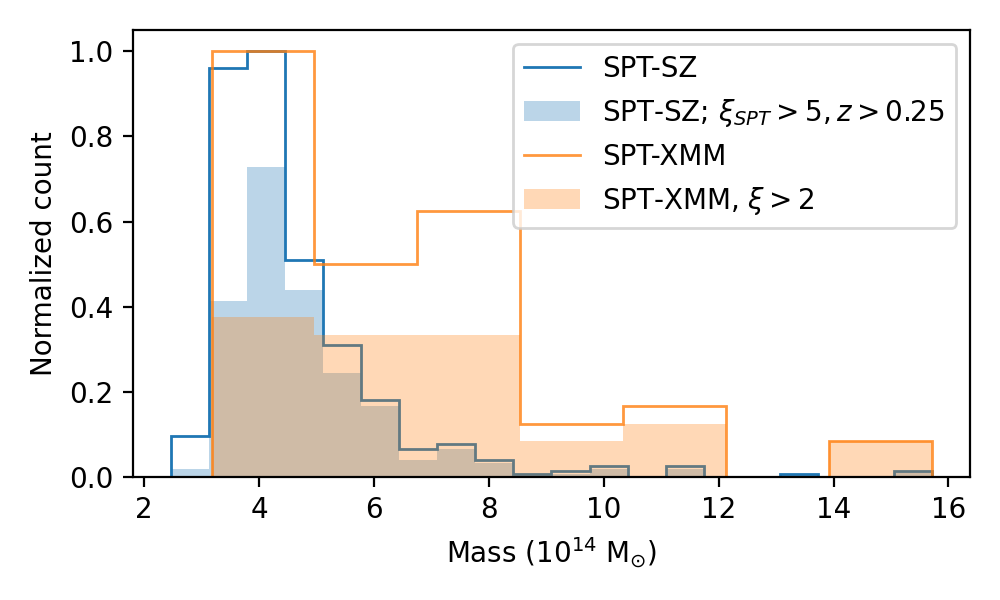}
            \caption{Mass and Redshift distributions of our (SPT-XMM) sample and the subset of clusters for which we have significant results ($\xi > 2$). We compare against the SPT-SZ sample and its subset used for cosmological constraints ($\xi_{\text{SPT}} > 5$ and $0.25 < z$). Distributions of subsets are normalized relative the their parent samples (retaining the same, respective, binning).}
            \label{fig:M_and_z_distributions}
        \end{figure}

        Here, we consider how the distribution of the clusters in our SPT-XMM sample compares to its ostensible parent distribution, the clusters in the SPT-SZ catalog. Although we may ultimately care about the distribution in the mass ($M_{500}$) - redshift plane, we describe their distributions against mass and redshift separately here. Namely, Figure~\ref{fig:M_and_z_distributions} shows how our SPT-XMM sample compares to the SPT-SZ sample (redshift and mass values are taken as those in \citet{bocquet2019}), and its subsample used for cosmological results applying the cuts , i.e. $z > 0.25$ and $\xi_{\text{SPT}} > 5$ resulting in 343 clusters \citep[e.g.][]{deHaan2016,bocquet2019}).

        In addition to Figure~\ref{fig:M_and_z_distributions}, we also calculate median redshifts and masses ($M_{500}$). The median redshifts for the full SPT-SZ sample and its cosmological subsample are 0.55 and 0.59, respectively. The median redshifts of our SPT-XMM sample and the 32 clusters which yielded significant results (amplitude spectra with at least one node of $\xi > 2$) are 0.44 and 0.42, respectively. The median masses of these last two samples are $5.63 \times 10^{14}$ M$_{\odot}$ and $5.89 \times 10^{14}$ M$_{\odot}$, respectively. The respective median redshifts of the full and cosmological SPT-SZ samples are $4.07 \times 10^{14}$ M$_{\odot}$ and $4.40 \times 10^{14}$ M$_{\odot}$.

    \subsection{Comparison of distributions}

        \begin{figure*}
            \centering
            \includegraphics[width=0.47\textwidth]{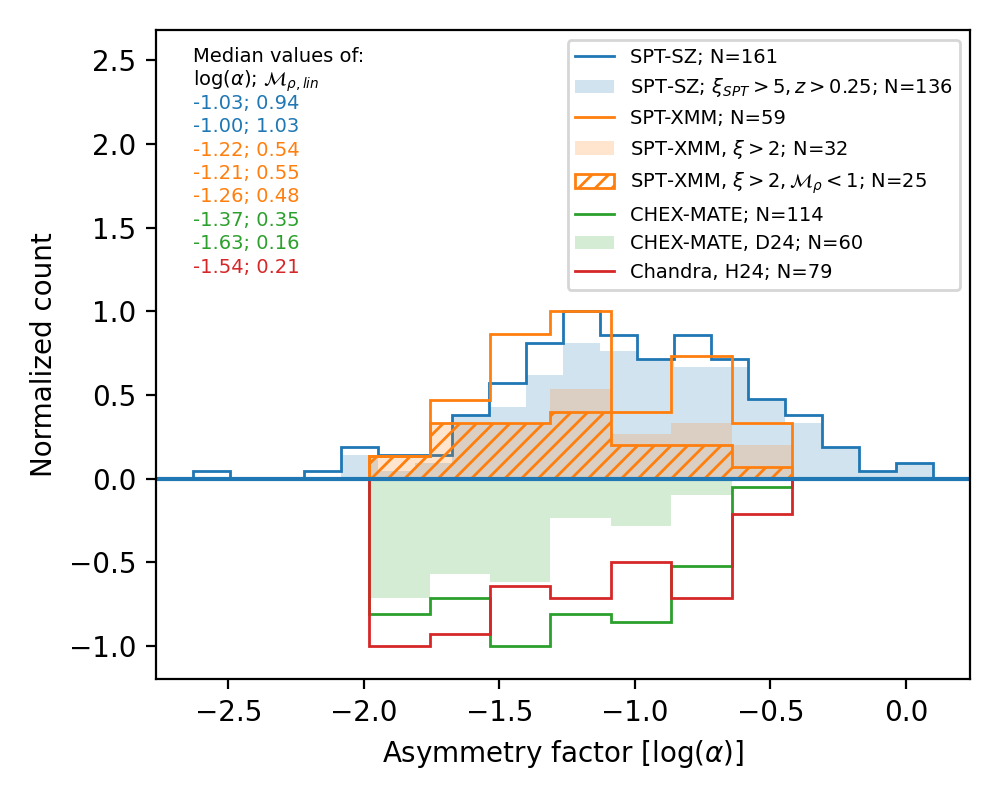}
            \includegraphics[width=0.47\textwidth]{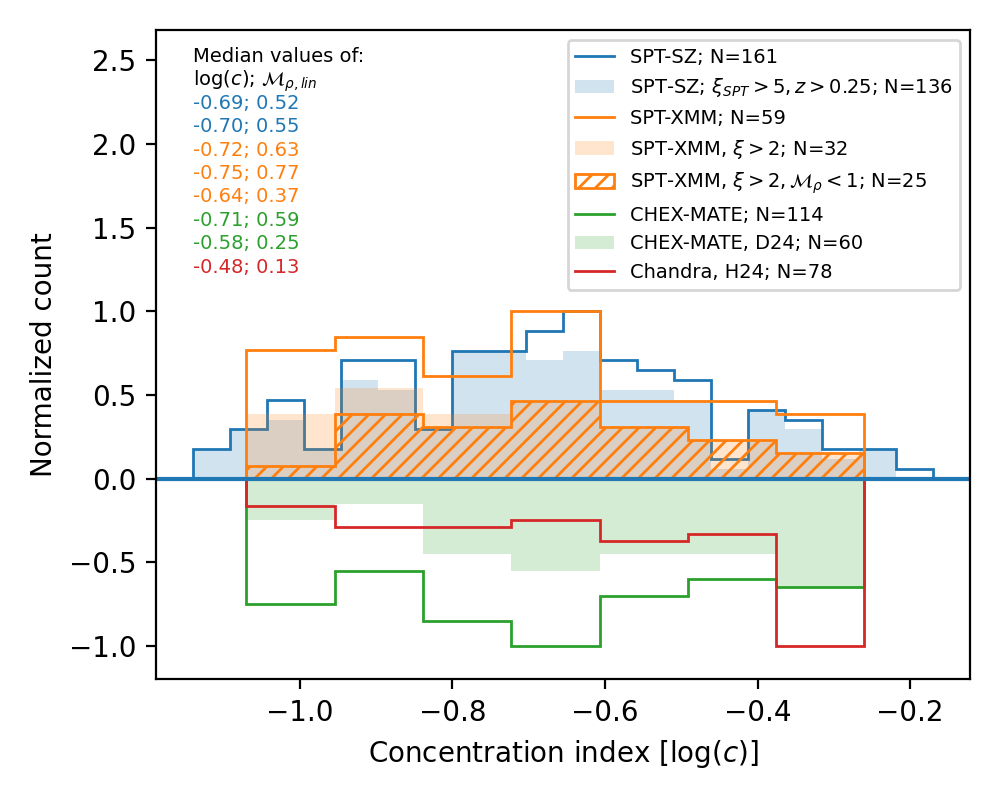}
            \includegraphics[width=0.47\textwidth]{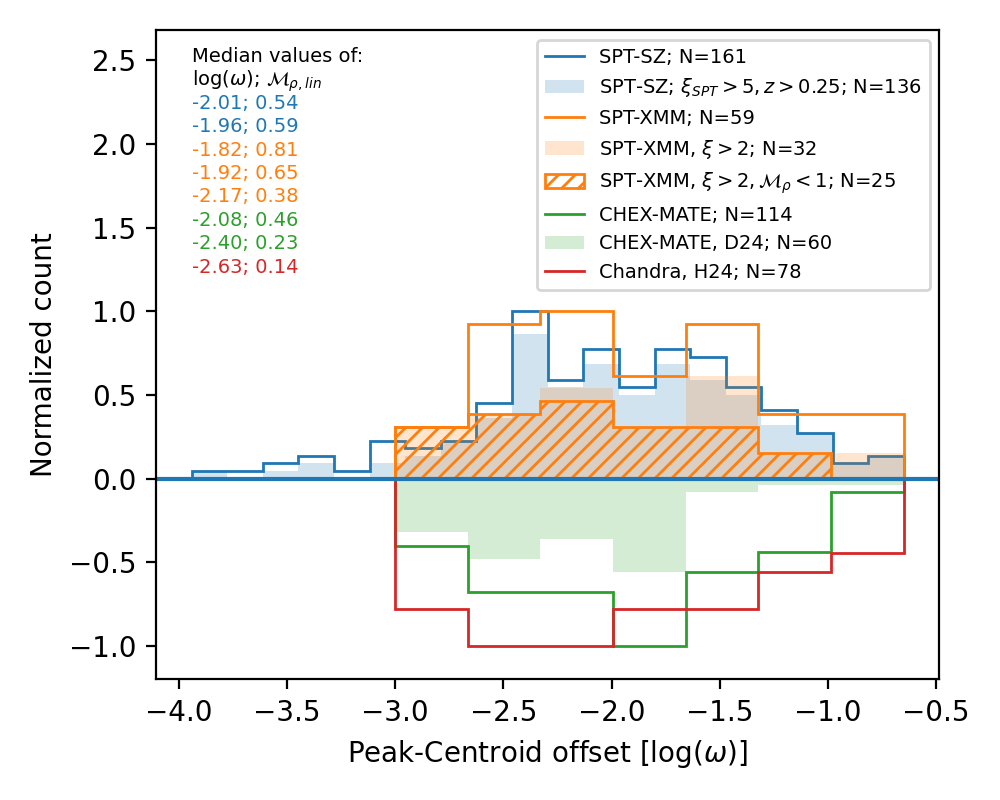}
            \includegraphics[width=0.47\textwidth]{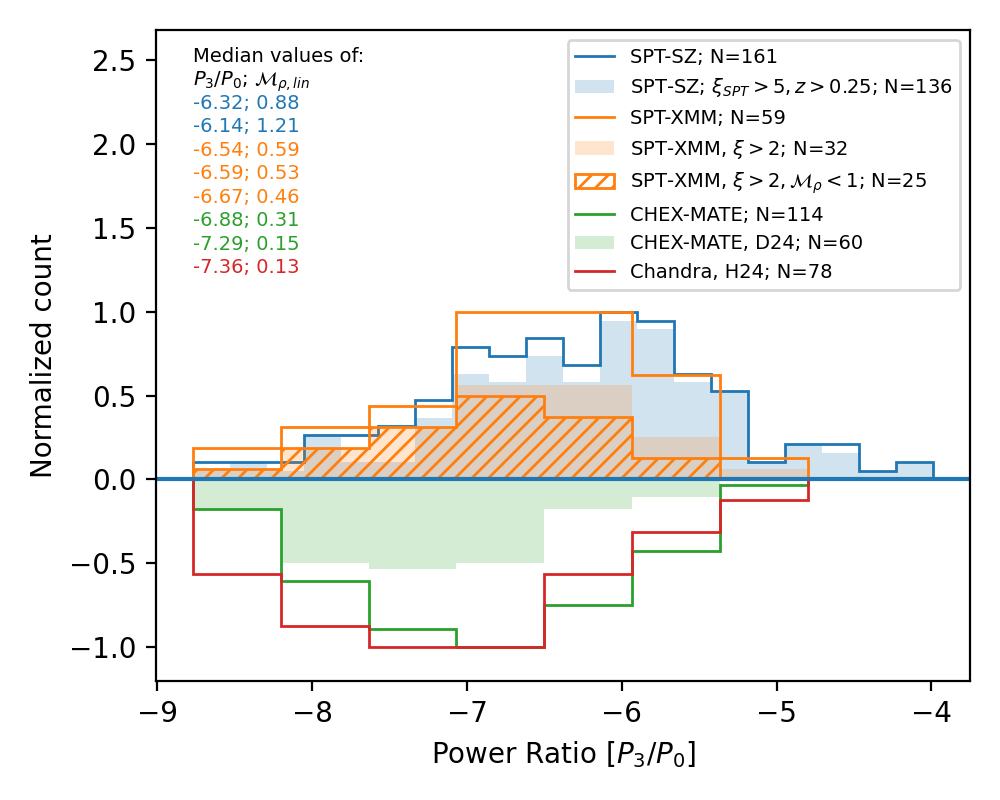}
            \includegraphics[width=0.47\textwidth]{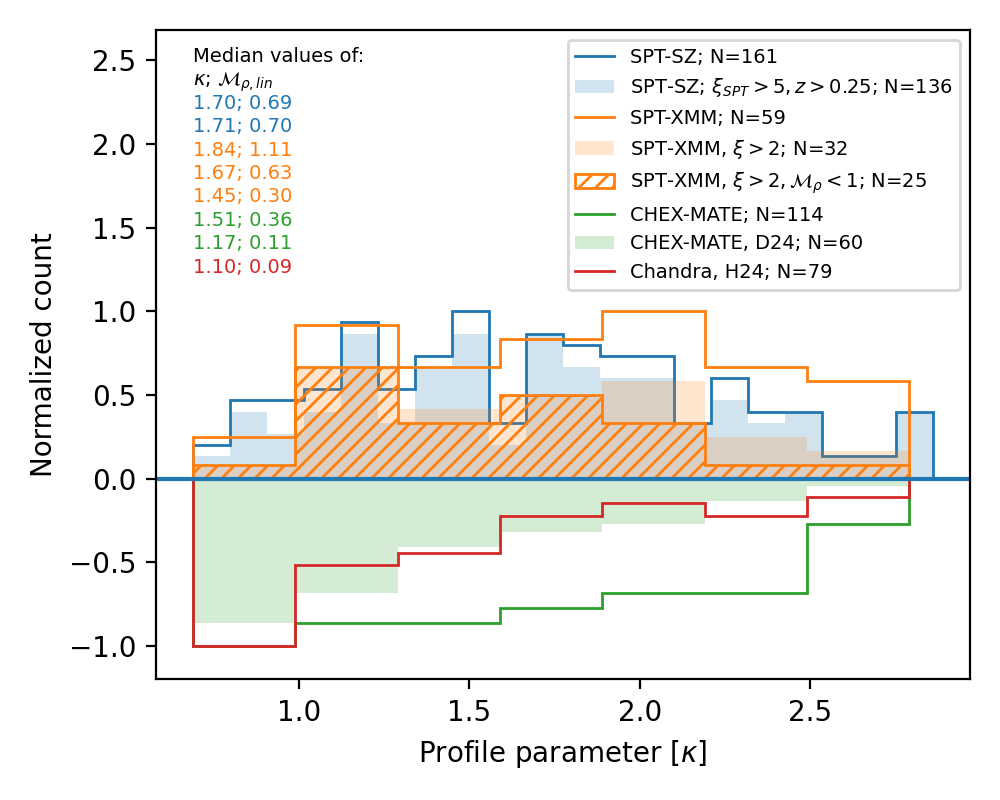}
            \includegraphics[width=0.47\textwidth]{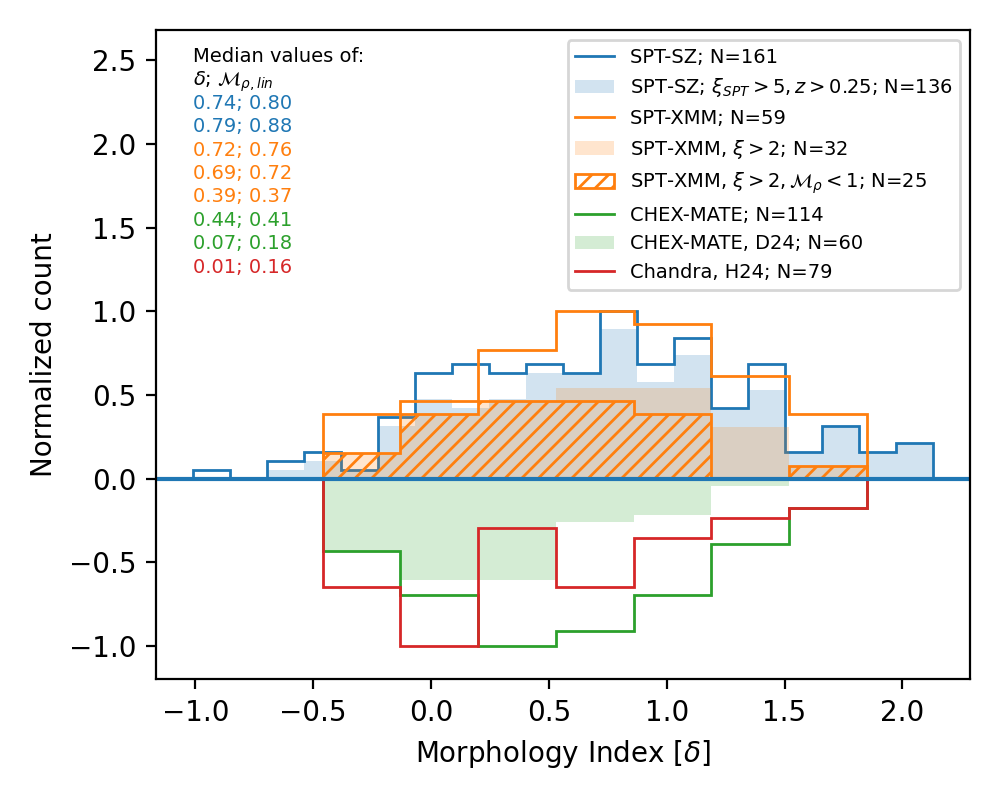}
            \caption{Distributions of the six dynamical parameters presented in \citet{yuan2022} for various samples (sign flipped for some samples to help with visibility). The number of clusters included, $N$, reflects that \citet{yuan2022} do not necessarily have dynamical parameters for all clusters in considered samples.
            Distributions of subsets are normalized relative the their parent samples (retaining the same, respective, binning). Text on the upper left of each panel indicates the median values of the dynamical parameter in question along with the median Mach number one would obtain from the linear coefficients presented in Table~\ref{tbl:A3D_M_coefficients}; the ordering (and coloring) matches that in the legend.}
            \label{fig:DynPar_distributions}
        \end{figure*}
   
        Analyzing our cluster distribution in mass and redshift space is appropriate to understand how representative our sample is of a proper SZ-selected sample. However, we also found in Section~\ref{sec:results} that density fluctuations (and thus our inferred turbulent velocities) did not correlate with mass (nor other quantities that should also correlate with mass) nor redshift. Indeed, this lack of correlation was also seen in \citet[][hereafter D24]{dupourque2024}. Thus, while we find the median redshifts of the samples in D24 and \citet[][hereafter H24]{heinrich2024} to be 0.19 and 0.17, respectively, we should not expect this to account for the difference in inferred Mach numbers between those studies, nor relative to our study (when limiting to our clusters with inferred subsonic turbulence, $z_{\text{med}} = 0.42$.) We do not consider a median mass comparison due to heterogeneous $M_{\delta}$ being reported, along with heterogeneous derivations of masses.

        Although our sample is not purely SZ selected, nor are the samples in H24 or D24 purely X-ray selected, we do, in fact see differences in the distribution of dynamical parameters between these samples, as shown in Figure~\ref{fig:DynPar_distributions}. In particular, we utilize the same parameters assessed in Section~\ref{sec:results}: $c$, $P_{3}/P_{0}$, $\alpha$, $\omega$, $\kappa$, and $\delta$, which (again) correspond to a concentration index,  power ratio, asymmetry factor, peak-centroid offset, profile parameter, and morphology index, respectively. In particular, our sample (even limiting to those clusters for which we infer subsonic turbulent velocities) tends to have more dynamically disturbed clusters than H24 or D24. Indeed, D24 made an explicit effort to omit merging systems (with their selection on the centroid offset parameter), which may be warranted in the hopes of properly tracing turbulence. Conversely, simulations have not selected on this same parameter, so one might expect a difference between simulations on this selection criterion alone.

        %We might also consider how these samples compare to a fuller sample of clusters. 
        %Unfortunately, we don't have a clear answer as we don't have access (yet) to the distributions of dynamical parameters for a truly SZ or X-ray selected sample. If it's the case that the subsample of SPT-SZ clusters with dynamical parameters is representative, then our sample (of 32 clusters) is roughly comparable, perhaps slightly more relaxed than either the SPT-SZ sample, or its subset (with restrictions of $z > 0.25$ and $\xi_{\text{SPT}} > 5$). Certainly our subsonic sample (25 clusters) is slightly more relaxed than either of the aforementioned SPT-SZ samples.

    %SPT-CLJ0234-5831, SPT-CLJ0343-5518, 

\input{SPTtable_GoodClusters}
\input{SPTtable_BadClusters}

%\end{landscape}

%% file: SPTtable_GoodClusters.tex
%\begin{longrotatetable}
%\startlongtable
%\begin{rotatetable}
\begin{deluxetable*}{ccccc ccccc cc} 
%\rotate 
 \centering  
\tablecaption{Cluster characteristics, observational properties, and inferred values \label{tbl:sample_good}}
\tabletypesize{\scriptsize}  % \small, \footnotesize, or \scriptsize
\tablehead{   
 \colhead{Cluster} & \colhead{$z^a$} & \colhead{$\theta_{500}$} & \colhead{$\xi_{\text{SPT}}^{a}$} & \colhead{$M_{500}^{b}$} & \colhead{Obs. ID} & \colhead{Exposures (ks)} & \colhead{Counts} & \colhead{$k_{\text{peak}}$} & \colhead{$A_{\rho,1}(k_{\text{peak}})$}  & \colhead{$\mathcal{M}_{\rho,1}$} & \colhead{$\xi_{A_{\rho,1}}$} \\[-0.25cm]
 \colhead{(SPT-CLJ)} & & \colhead{(arcmin)} & & \colhead{($10^{14}$ M$_{\odot}$)} &  & \colhead{MOS1;MOS2;PN} & \colhead{MOS1;MOS2;PN} & ($R_{500}^{-1}$) & & & (max) 
 } 
\startdata 
2248-4431 & 0.35 & 4.96 & 42.36 & 13.05 & 0504630101 & 25.70;26.60;21.90 & 14127;14012;36984 & 3.97 & $0.10 \pm 0.05$ & $0.44 \pm 0.22$ & 12.09 \\ 
0658-5556 & 0.29 & 5.71 & 39.05 & 12.70 & 0112980201 & 22.20;22.20;18.00 & 12730;12380;31034 & 4.57 & $0.28 \pm 0.05$ & $1.29 \pm 0.21$  & 15.91 \\ 
0549-6205 & 0.37 & 4.30 & 25.81 & 9.66 & 0656201301 & 13.40;13.10;9.90 & 4223;3908;10405 & 3.44 & $0.12 \pm 0.05$ & $0.53 \pm 0.22$ & 7.78 \\ 
  &   &   &   &   & 0827050701 & 39.50;39.70;37.90 & 12057;11211;37560 &   &   &   \\ 
0232-4421 & 0.28 & 5.33 & 23.96 & 9.45 & 0042340301 & 11.60;12.10;6.80 & 4414;4757;9652 & 2.63 & $0.16 \pm 0.02$ & $0.65 \pm 0.08$ & 9.51 \\ 
  &   &   &   &   & 0827350201 & 24.60;25.70;18.00 & 9403;9393;26991 &   &   &   \\ 
0638-5358 & 0.23 & 6.26 & 22.69 & 9.42 & 0650860101 & 24.60;31.70;7.60 & 11177;14200;13067 & 1.71 & $0.18 \pm 0.02$ & $0.64 \pm 0.06$ & 11.49 \\ 
0438-5419 & 0.42 & 3.77 & 22.88 & 8.68 & 0656201601 & 18.00;18.00;13.50 & 3002;2907;7331 & 3.01 & $0.14 \pm 0.05$ & $0.57 \pm 0.21$ & 9.21 \\ 
  &   &   &   &   & 0827360501 & 37.80;40.30;33.60 & 6230;6386;18737 &   &   &   \\ 
2031-4037 & 0.34 & 4.31 & 17.52 & 7.95 & 0690170501 & 2.50;2.50;0.80 & 352;368;432 & 1.00 & $0.08 \pm 0.04$ & $0.26 \pm 0.11$ & 2.30 \\ 
  &   &   &   &   & 0690170701 & 10.30;10.10;8.60 & 1565;1522;4800 &   &   &   \\ 
2106-5844 & 1.13 & 1.80 & 22.22 & 7.14 & 0744400101 & 41.50;46.30;19.30 & 1146;1055;1901 & 1.00 & $0.22 \pm 0.04$ & $0.70 \pm 0.12$ & 5.75 \\ 
  &   &   &   &   & 0763670301 & 26.80;27.70;18.40 & 688;746;1729 &   &   &   \\ 
2337-5942 & 0.77 & 2.28 & 20.35 & 7.05 & 0604010201 & 18.20;19.70;10.20 & 740;737;1462 & 1.00 & $0.13 \pm 0.05$ & $0.42 \pm 0.17$ & 2.49 \\ 
0304-4401 & 0.46 & 3.27 & 15.69 & 6.98 & 0700182201 & 16.90;16.80;13.00 & 1570;1457;3926 & 2.62 & $0.34 \pm 0.05$ & $1.37 \pm 0.19$  & 10.06 \\ 
2023-5535 & 0.23 & 5.53 & 13.63 & 6.49 & 0841951701 & 13.90;14.00;11.40 & 2338;2402;6837 & 1.00 & $0.19 \pm 0.02$ & $0.59 \pm 0.06$ & 9.75 \\ 
0243-4833 & 0.50 & 2.97 & 13.90 & 6.26 & 0672090501 & 10.40;10.20;5.30 & 772;772;1651 & 1.00 & $0.09 \pm 0.03$ & $0.29 \pm 0.09$ & 3.23 \\ 
  &   &   &   &   & 0723780801 & 12.70;11.60;3.70 & 987;980;1063 &   &   &   \\ 
2138-6008 & 0.32 & 4.14 & 12.64 & 6.10 & 0674490201 & 13.10;14.40;9.80 & 1260;1340;2822 & 1.00 & $0.07 \pm 0.03$ & $0.22 \pm 0.08$ & 2.70 \\ 
0114-4123 & 0.38 & 3.57 & 11.43 & 5.86 & 0724770901 & 12.40;12.80;7.40 & 926;973;2034 & 1.00 & $0.07 \pm 0.03$ & $0.21 \pm 0.08$ & 2.63 \\ 
0014-3022 & 0.12 & 9.20 & 18.29 & 5.43 & 0042340101 & 13.90;14.20;10.90 & 3337;3499;7740 & 14.40 & $0.28 \pm 0.13$ & $1.70 \pm 0.79$ & 19.47 \\ 
  &   &   &   &   & 0743850101 & 96.50;96.60;82.80 & 22015;21716;58130 &   &   &   \\ 
0559-5249 & 0.61 & 2.39 & 10.64 & 5.03 & 0604010301 & 18.30;18.20;13.50 & 415;391;1059 & 1.00 & $0.18 \pm 0.04$ & $0.57 \pm 0.14$ & 4.02 \\ 
2341-5119 & 1.00 & 1.71 & 12.49 & 4.94 & 0744400401 & 74.30;84.20;47.30 & 1189;1294;2869 & 1.00 & $0.12 \pm 0.03$ & $0.39 \pm 0.10$ & 3.87 \\ 
  &   &   &   &   & 0763670201 & 31.10;35.20;16.10 & 503;527;990 &   &   &   \\ 
2146-4633 & 0.93 & 1.79 & 9.67 & 4.89 & 0744400501 & 94.10;97.90;70.50 & 1209;1148;3466 & 1.43 & $0.16 \pm 0.05$ & $0.56 \pm 0.16$ & 3.73 \\ 
  &   &   &   &   & 0744401301 & 71.50;75.30;44.50 & 883;917;2277 &   &   &   \\ 
0240-5946 & 0.40 & 3.22 & 8.84 & 4.85 & 0674490101 & 14.30;14.20;7.90 & 779;759;1381 & 1.00 & $0.17 \pm 0.06$ & $0.54 \pm 0.20$ & 2.73 \\ 
2032-5627 & 0.28 & 4.24 & 8.61 & 4.77 & 0674490401 & 25.10;25.80;19.30 & 3162;3538;8248 & 3.39 & $0.56 \pm 0.04$ & $2.40 \pm 0.17$ & 14.62 \\ 
2124-6124 & 0.44 & 2.94 & 8.50 & 4.60 & 0674490701 & 14.10;14.70;7.90 & 434;487;862 & 1.00 & $0.14 \pm 0.07$ & $0.46 \pm 0.22$ & 2.06 \\ 
0225-4155 & 0.22 & 5.02 & 6.92 & 4.33 & 0692933401 & 12.50;12.20;10.90 & 3862;3643;11651 & 4.01 & $0.16 \pm 0.01$ & $0.74 \pm 0.05$ & 17.02 \\ 
  &   &   &   &   & 0803550101 & 64.20;68.40;50.70 & 19341;21865;42391 &   &   &   \\ 
2017-6258 & 0.53 & 2.46 & 6.32 & 4.03 & 0674491501 & 25.90;25.80;20.80 & 328;273;833 & 1.00 & $0.41 \pm 0.11$ & $1.30 \pm 0.34$ & 3.80 \\ 
0344-5452 & 1.00 & 1.58 & 7.98 & 3.89 & 0675010701 & 49.50;49.70;43.00 & 303;248;1044 & 1.00 & $0.16 \pm 0.08$ & $0.51 \pm 0.24$ & 2.12 \\ 
0254-5857 & 0.44 & 2.78 & 14.13 & 3.86 & 0656200301 & 11.90;13.30;6.80 & 1081;1368;2116 & 1.49 & $0.15 \pm 0.01$ & $0.53 \pm 0.05$ & 10.84 \\ 
  &   &   &   &   & 0674380301 & 45.90;47.20;38.90 & 4493;4383;12464 &   &   &   \\ 
0354-5904 & 0.41 & 2.92 & 6.42 & 3.83 & 0724770501 & 14.80;16.30;9.10 & 333;554;968 & 1.00 & $0.33 \pm 0.05$ & $1.06 \pm 0.17$ & 6.27 \\ 
0317-5935 & 0.47 & 2.61 & 6.26 & 3.73 & 0674490501 & 8.10;10.90;1.90 & 232;270;204 & 1.00 & $0.13 \pm 0.05$ & $0.41 \pm 0.17$ & 2.44 \\ 
  &   &   &   &   & 0724770401 & 15.00;15.10;7.30 & 567;489;921 &   &   &   \\ 
0233-5819 & 0.66 & 2.05 & 6.55 & 3.70 & 0675010601 & 49.70;50.90;38.30 & 754;736;2183 & 1.00 & $0.10 \pm 0.03$ & $0.31 \pm 0.11$ & 2.96 \\ 
0403-5719 & 0.46 & 2.60 & 5.86 & 3.52 & 0674491201 & 18.60;20.00;10.10 & 994;1104;1893 & 1.44 & $0.15 \pm 0.05$ & $0.52 \pm 0.18$ & 3.35 \\ 
0522-4818 & 0.29 & 3.67 & 4.82 & 3.37 & 0303820101 & 11.60;15.30;3.10 & 863;957;680 & 1.00 & $0.14 \pm 0.03$ & $0.45 \pm 0.11$ & 4.19 \\ 
2056-5459 & 0.72 & 1.87 & 6.07 & 3.36 & 0675010901 & 40.70;39.80;36.00 & 371;396;1167 & 1.00 & $0.42 \pm 0.10$ & $1.33 \pm 0.32$ & 4.14 \\ 
2011-5725 & 0.28 & 3.77 & 5.34 & 3.35 & 0744390401 & 17.20;17.70;10.50 & 739;799;1036 & 1.74 & $0.16 \pm 0.05$ & $0.59 \pm 0.20$ & 5.18 
\enddata 
\tablecomments{ {\footnotesize Properties of clusters for which at least one node in $A_{\rho,1}$ has a SNR of $\xi > 2$. $^{a}$Values taken from \citet{bocquet2019} and for SPT-CLJ0014-3022 from \citet{plagge2010}. $\xi_{\text{SPT}}$ refers to the detection significance of the cluster from SPT data \citep{bleem2015}. $^{b}$Values taken from \citet{bulbul2019}. $\theta_{500}$ is inferred from $M_{500}$ and our assumed cosmology. $\xi_{\rho,1}$ refers to the maximum significance of nodes within the amplitude spectrum $A_{\rho,1}$.} } 
\end{deluxetable*} 
%\end{rotatetable}
%\end{longtable}
%\end{longrotatetable}

%% file: SPTtable_BadClusters.tex
\begin{deluxetable*}{ccccc ccc} 
%\rotate 
 \centering  
\tablecaption{Cluster characteristics and observational properties for non-detections \label{tbl:sample_bad}}
\tabletypesize{\small}  % \small, \footnotesize, or \scriptsize
\tablehead{   
 & \colhead{$z^a$} & \colhead{$\theta_{500}$} & \colhead{$\xi_{\text{SPT}}^a$} & \colhead{$M_{500}^b$} & \colhead{Obs. ID} & \colhead{Exposures (ks)} & \colhead{Counts} \\[-0.25cm] 
 \colhead{Cluster} & & \colhead{(arcmin)} & & \colhead{$10^{14}$ M$_{\odot}$} &  & \colhead{MOS1;MOS2;PN} & \colhead{MOS1;MOS2;PN}
 } 
\startdata 
SPT-CLJ0615-5746 & 0.97 & 2.11 & 26.42 & 8.69 & 0658200101 & 12.70;13.40;5.20 & 641;643;663 \\ 
SPT-CLJ0234-5831 & 0.42 & 3.45 & 14.66 & 6.70 & 0674491001 & 12.70;13.80;9.10 & 1485;1580;3558 \\ 
SPT-CLJ2131-4019 & 0.45 & 3.21 & 12.51 & 6.25 & 0724770601 & 12.70;12.90;6.20 & 1288;1395;2417 \\ 
SPT-CLJ0417-4748 & 0.58 & 2.66 & 14.24 & 6.22 & 0700182401 & 22.10;23.80;15.30 & 1646;1754;3981 \\ 
SPT-CLJ0516-5430 & 0.29 & 4.44 & 12.41 & 5.96 & 0042340701 & 5.00;5.00;0.80 & 1006;1096;482 \\ 
  &   &   &   &   & 0205330301 & 10.40;10.70;8.10 & 2166;2210;5323 \\ 
  &   &   &   &   & 0692934301 & 27.50;27.40;23.60 & 5452;5513;14735 \\ 
SPT-CLJ2145-5644 & 0.48 & 2.98 & 12.60 & 5.82 & 0674491301 & 10.30;10.70;6.40 & 619;666;1221 \\ 
SPT-CLJ0510-4519 & 0.20 & 5.97 & 9.50 & 5.73 & 0692933001 & 13.00;13.10;11.10 & 4007;3975;11859 \\ 
SPT-CLJ0205-5829 & 1.32 & 1.39 & 10.40 & 4.37 & 0675010101 & 57.00;57.90;46.70 & 472;412;1208 \\ 
  &   &   &   &   & 0803050201 & 10.50;12.40;6.00 & 82;74;150 \\ 
SPT-CLJ2130-6458 & 0.31 & 3.78 & 7.63 & 4.33 & 0692900101 & 6.30;8.20;4.10 & 403;489;938 \\ 
SPT-CLJ0254-6051 & 0.44 & 3.31 & 6.55 & 6.52 & 0692900201 & 16.20;15.70;12.20 & 316;334;1027 \\ 
SPT-CLJ0217-5245 & 0.34 & 3.43 & 6.46 & 4.01 & 0652951401 & 9.30;14.70;3.80 & 332;470;448 \\ 
SPT-CLJ2022-6323 & 0.38 & 3.09 & 6.51 & 3.80 & 0674490601 & 14.70;14.40;5.80 & 290;201;373 \\ 
SPT-CLJ2200-6245$^c$ & 0.39 & 3.02 & 0.00 & 3.79 & 0674490801 & 9.60;10.70;6.20 & 180;140;343 \\ 
  &   &   &   &   & 0724771001 & Not used & Not used \\ 
SPT-CLJ0343-5518 & 0.55 & 2.29 & 6.01 & 3.52 & 0724770801 & 18.10;18.00;11.80 & 252;265;635 \\ 
SPT-CLJ0230-6028 & 0.68 & 1.95 & 6.01 & 3.43 & 0675010401 & 19.50;25.40;11.20 & 295;412;705 \\ 
SPT-CLJ2030-5638 & 0.39 & 2.90 & 5.50 & 3.35 & 0724770201 & 21.10;21.10;17.10 & 391;398;1133 \\ 
SPT-CLJ2040-4451 & 1.48 & 1.19 & 6.72 & 3.31 & 0723290101 & 76.30;76.10;72.90 & 280;282;1019 \\ 
SPT-CLJ0406-5455 & 0.74 & 1.82 & 5.91 & 3.28 & 0675010501 & 54.20;56.10;40.00 & 426;354;1218 \\ 
SPT-CLJ2136-6307 & 0.93 & 1.56 & 6.24 & 3.24 & 0675010301 & 57.30;60.80;50.30 & 429;417;1164 \\ 
SPT-CLJ2040-5725 & 0.93 & 1.56 & 6.24 & 3.23 & 0675010201 & 75.70;77.30;68.40 & 783;600;2113 \\ 
SPT-CLJ0231-5403 & 0.59 & 2.10 & 5.22 & 3.18 & 0204530101 & 17.30;22.00;4.50 & 195;320;146 \\ 
SPT-CLJ0257-5732 & 0.43 & 2.64 & 5.04 & 3.15 & 0674491101 & 27.60;28.10;23.10 & 180;125;530 \\ 
SPT-CLJ0611-5938 & 0.39 & 2.84 & 4.74 & 3.13 & 0658201101 & 13.10;13.40;6.30 & 367;302;616 \\ 
SPT-CLJ2109-4626 & 0.97 & 1.43 & 4.65 & 2.68 & 0694380101 & 53.10;56.10;43.30 & 224;159;593 
\enddata 
\tablecomments{Properties of clusters for which amplitude spectra were produced but for which the SNR threshold $\xi > 2$ was not met. $^{a}$Values taken from \citet{bocquet2019} and for SPT-CLJ0014-3022 from \citet{plagge2010}. $\xi_{\text{SPT}}$ refers to the detection significance of the cluster from SPT data \citep{bleem2015}. $^{b}$Values taken from \citet{bulbul2019}. $\theta_{500}$ is inferred from $M_{500}$ and our assumed cosmology. $^c$Listed with this moniker in \citet{bulbul2019}, it is more commonly found with the moniker SPT-CLJ2159-6244.} 
\end{deluxetable*}

%% file: Appendix_Substructure.tex
            We employ an algorithm to identify substructure and its extent using net rate images smoothed with three different Gaussian kernels. For each EPIC camera and each energy band, we apply the three smoothing kernels, indexed with $k$. For each smoothing kernal and within each annulus, indexed with $j$, of our radial profile we calculate pixel means, $\bar{p}_{j,k}$ and the RMS within that annulus (for that smoothing), i.e. $\sigma_{j,k}$. For a normal distribution, and a given number of pixels, we can infer that for some multiplicative factor, $f$ of the RMS, we will not expect to find any pixels with value $p > \bar{p} + f \sigma$. Within the context of our annuli and smoothings, we flag pixels with $p_{j,k} > \bar{p}_{j,k} + f_{j} \sigma_{j,k}$. A mask could then be produced per smoothing of each EPIC camera.

            For a given smoothing, we expect similar substructure to be flagged across the EPIC cameras. However, visual inspection showed that what was masked in one camera may not be masked (or with many fewer pixels masked) in another camera. To ameliorate this, we introduced another factor, per smoothing kernal, $g_{k}$ such that we flag pixels $p_{j,k} > \bar{p}_{j,k} + g_{k} f_{j} \sigma_{j,k}$. The values of the adjustments are modest, where $0.5 < g < 2$ yields visual consistency across the cameras. Visual consistency was not rigorously defined, but approximately corresponds to the number of pixels for a given substructure matching within a factor of 2 across all cameras. Initially, a given $g_k$ value which may have produced (visual) consistency for one cluster did not produce (visual) consistency for another. After some iteration, we found values of $g_k$, independent of cluster, which produced consistency.
            %For each of these images, for a given annulus $i$, pixel means, $\bar{p}_i$, and Gaussian widths, $\sigma_{j}$ are determined within annuli of $10^{\prime\prime}$ width. Initially, a threshold, expressed as a factor $f_{i,j}$ is chosen per annulus and per smoothing such that a pixel is not expected to surpass that threshold for a Gaussian distribution. That is, we flag pixels with value $p > f_{i,j} \sigma_{j}$. However, we found that this threshold did not produce consistent flagged regions across the EPIC cameras (MOS1, MOS2, and pn). To improve this consistency, we added a multiplicative factor, $g_{j}$, per smoothing kernel, such that we then flag pixels with $p > g_j f_{i,j} \sigma_{i}$. We find that factors of $0.5 < g < 2$ were able to create greater consistency across EPIC cameras. %In particular, we desired that regions were not spuriously identified as substructure within $R_{500}$. This results in a conservative masking which may still leave substructure unmasked.            
            This flagging yielded binary masks per EPIC camera (MOS1, MOS2, and pn) and each energy band (400-1250 eV) and (2000-5000 eV). 
            For each cluster, we stacked the binary masks from each EPIC camera and energy band, gently smoothed the stacked mask, and employed another threshold to obtain a merged (binary) mask which closely matched the individual masks.
            
            \begin{figure}
                \centering
                \includegraphics[width=0.4\textwidth]{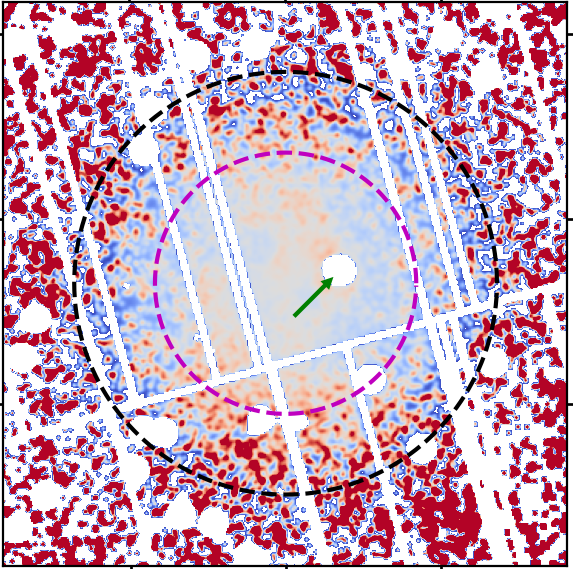}
                \includegraphics[width=0.4\textwidth]{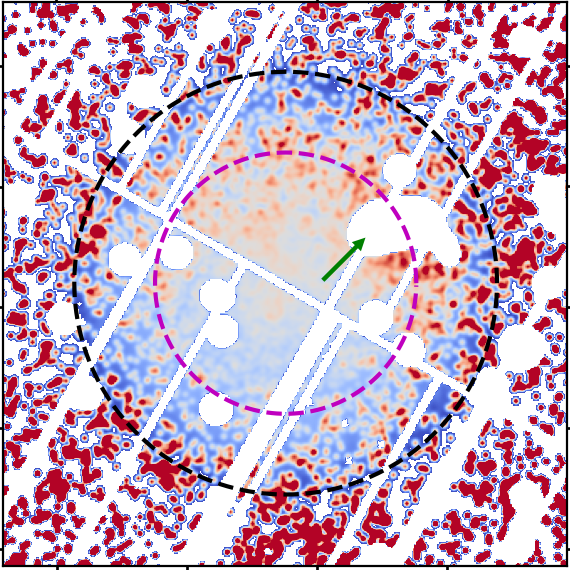}
                \caption{The $\delta S/\bar{S}_{\text{ICM}}$ image for SPT-CLJ0658-5556 (top) and SPT-CLJ0014-3022 (bottom) as seen with the pn camera (400-1250 eV). The substructures masked are indicated with green arrows; other masked regions are from point sources and chip gaps.}
                \label{fig:bullet_cluster}
            \end{figure}
            %In SPT-CLJ0658-5556, the nearly circular masked region inside the green circle was determined from the substructure finding algorithm indicated in the text. In SPT-CLJ0014-3022

\begin{figure*}
                \centering
                \includegraphics[width=0.47\textwidth]{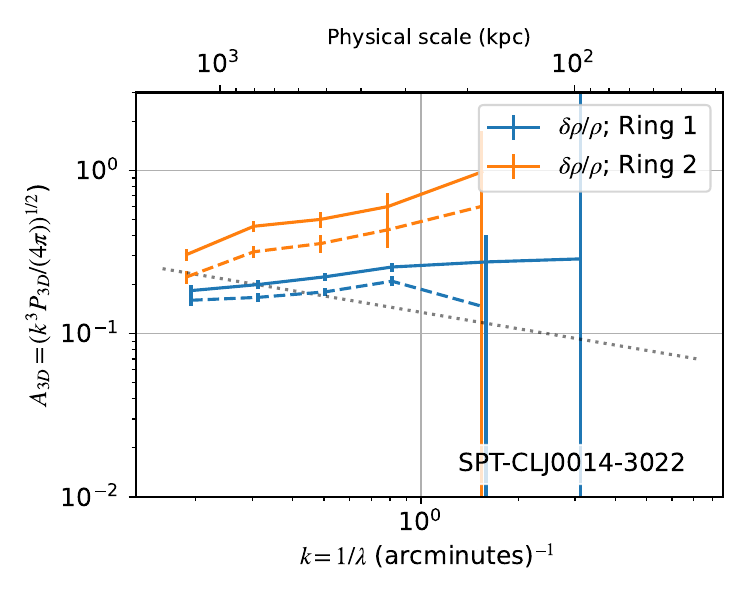}
                \includegraphics[width=0.47\textwidth]{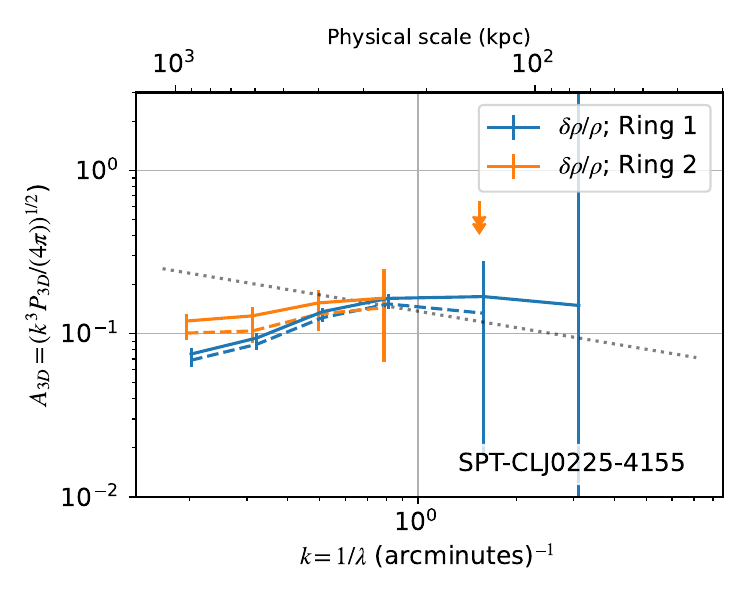}
                \includegraphics[width=0.47\textwidth]{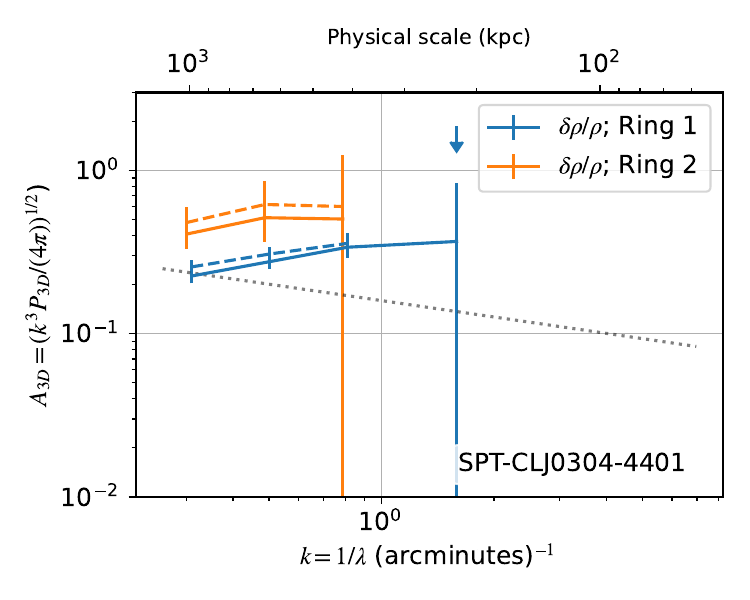}
                \includegraphics[width=0.47\textwidth]{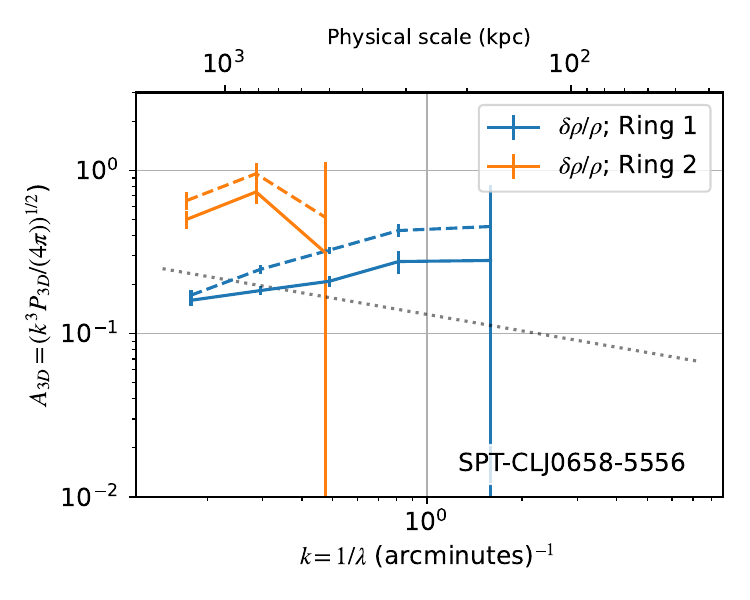}
                \caption{The resultant amplitude spectra, $A_{\rho}$, for clusters with substructure masked (solid lines) and unmasked (dashed lines). The clusters are SPT-CLJ0014-3022 (top left), SPT-CLJ0225-4155 (top right), SPT-CLJ0304-4401 (bottom left), and SPT-CLJ0658-5556 (bottom right). Arrows indicate a $3\sigma$ upper limit.}
                \label{fig:A3D_w-wo_ss}
            \end{figure*}

            Figure~\ref{fig:bullet_cluster} shows the normalized residuals, $\delta S / \bar{S}_{\text{ICM}}$, for SPT-CLJ0658-5556, with the substructure masking algorithm masking solely the bullet (and not the bow shock). In the case of the bullet cluster, masking the substructure (the bullet) reduces the recovered fluctuations as seen in the amplitude spectra (Figure~\ref{fig:A3D_w-wo_ss}). However, for some clusters (e.g. SPT-CLJ0014-3022 and SPT-CLJ0225-4155), the amplitudes can increase. Much as in \citet{romero2024}, changes in the surface brightness profile modeling, such as masking, which induce a steeper profiles (thus smaller $\bar{S}_{\text{ICM}}$ values) can ultimately yield larger fluctuations ($\delta S / \bar{S}_{\text{ICM}} $). Another effect is that, for a fixed a $P_{\text{2D}}$, the deprojection will produce larger values of $P_{\text{3D}}$ , and thus $A_{\text{3D}}$, relative to its counterpart from a surface brightness profile with a shallower slope. 

%% file: Appendix_Spectra.tex
%\textcolor{red}{I will produce figures of the amplitude spectra of clusters which have at least one spectral point with significance above 9.}

\begin{figure*}
        \centering
        \includegraphics[width=0.97\textwidth]{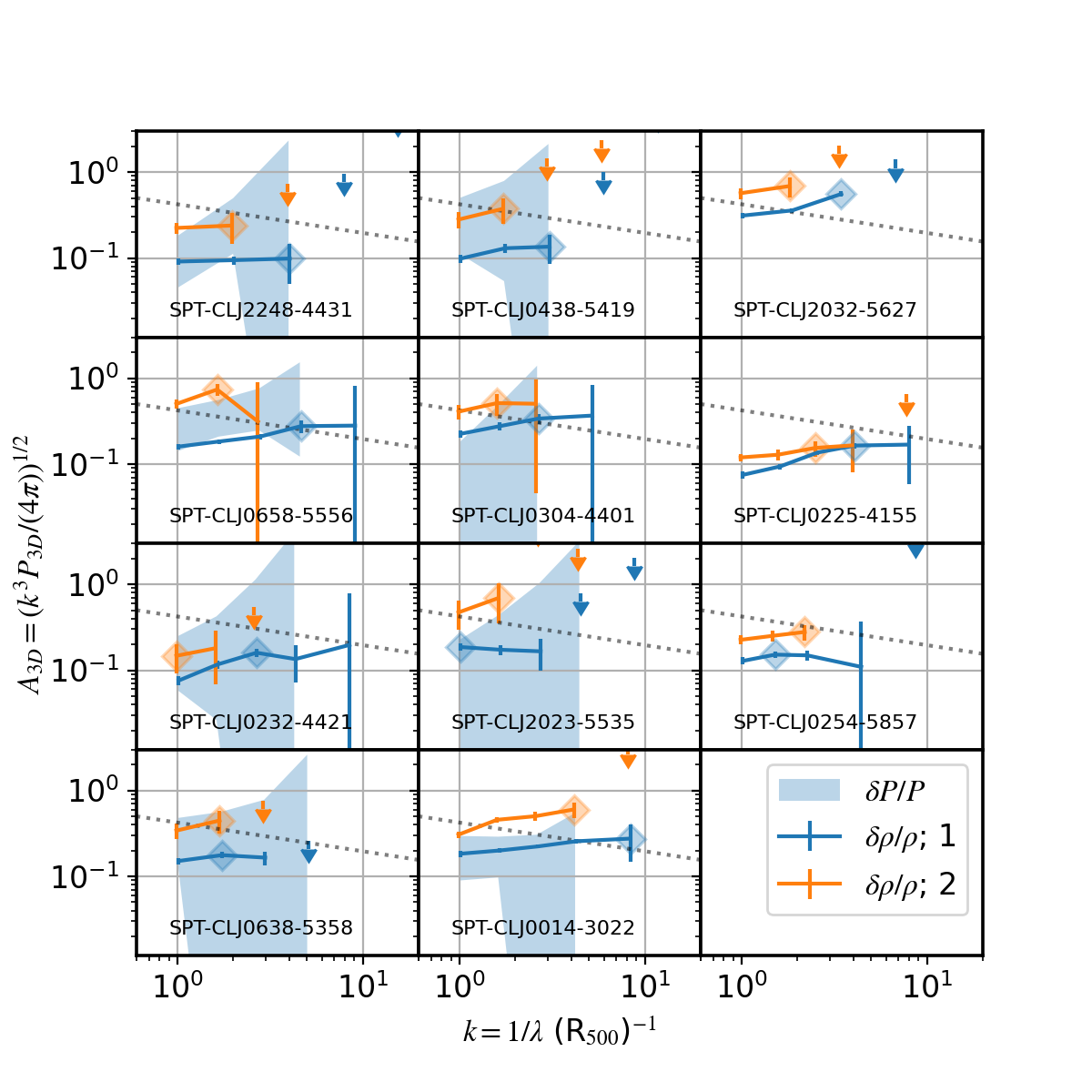}
        \caption{Amplitude spectra of density fluctuations (lines; blue corresponds to Ring 1 and orange to Ring 2) and pressure fluctuations in Ring 1 (shaded region, if significant) for clusters with $\xi > 9$ for at least one node of $A_{\rho}$ in Ring 1. Diamonds indicate which node is taken as the peak (of nodes with $\xi > 2$). Arrows indicate a $3\sigma$ upper limit.}
        \label{fig:A3Ds}
    \end{figure*}

    In Figure~\ref{fig:A3Ds} we present the amplitude spectra of those clusters for which $A_{\rho}$ in Ring 1 had at least one node with significance $\xi > 9$. This is an arbitrary choice to showcase a handful of clusters with the best data (especially spectra with three or more nodes of $\xi > 2$). Without clear observations of the spectral cascade, i.e. significant constraints at scales smaller than the observed peaks (with $\xi_{A_{\rho}} > 2$), we are limited in how well we can infer the injection scales. To the extent that a drop-off at larger scales than the injection scale is expected \citep[e.g.][]{Gaspari2013_PS}, the relatively flat spectra (e.g. that of Ring 1 in SPT-CLJ2248-4431 or SPT-CLJ0014-3022, for which many nodes have $\xi > 5$) suggest multiple injection scales. 
    
    This potential has been noted in other works \citep[e.g.][]{romero2023,dupourque2023,romero2024}, and in the case of SPT-CLJ0014-3022 (that is, Abell 2744), we know it is a multiple-merger system. Notwithstanding issues of substructure masking (discussed in Appendix~\ref{sec:substructure_masking}), it is not surprising to find multiple injection scales. \citet{gomez2012} find a bimodal galaxy distribution in SPT-CLJ2248-4431 (Abell S1063) and infer that it is in a merging state, while the X-ray distribution does not reveal such bimodality. \citet{shitanishi2018} classify SPT-CLJ2248-4431 as a non cool-core cluster and there is no substantial substructure in the X-ray images, including work by \citet{olivares2023} who found no evidence of X-ray cavities in \textit{Chandra} images of SPT-CLJ2248-4431. 

    \begin{figure}
        \centering
        \includegraphics[width=0.47\textwidth]{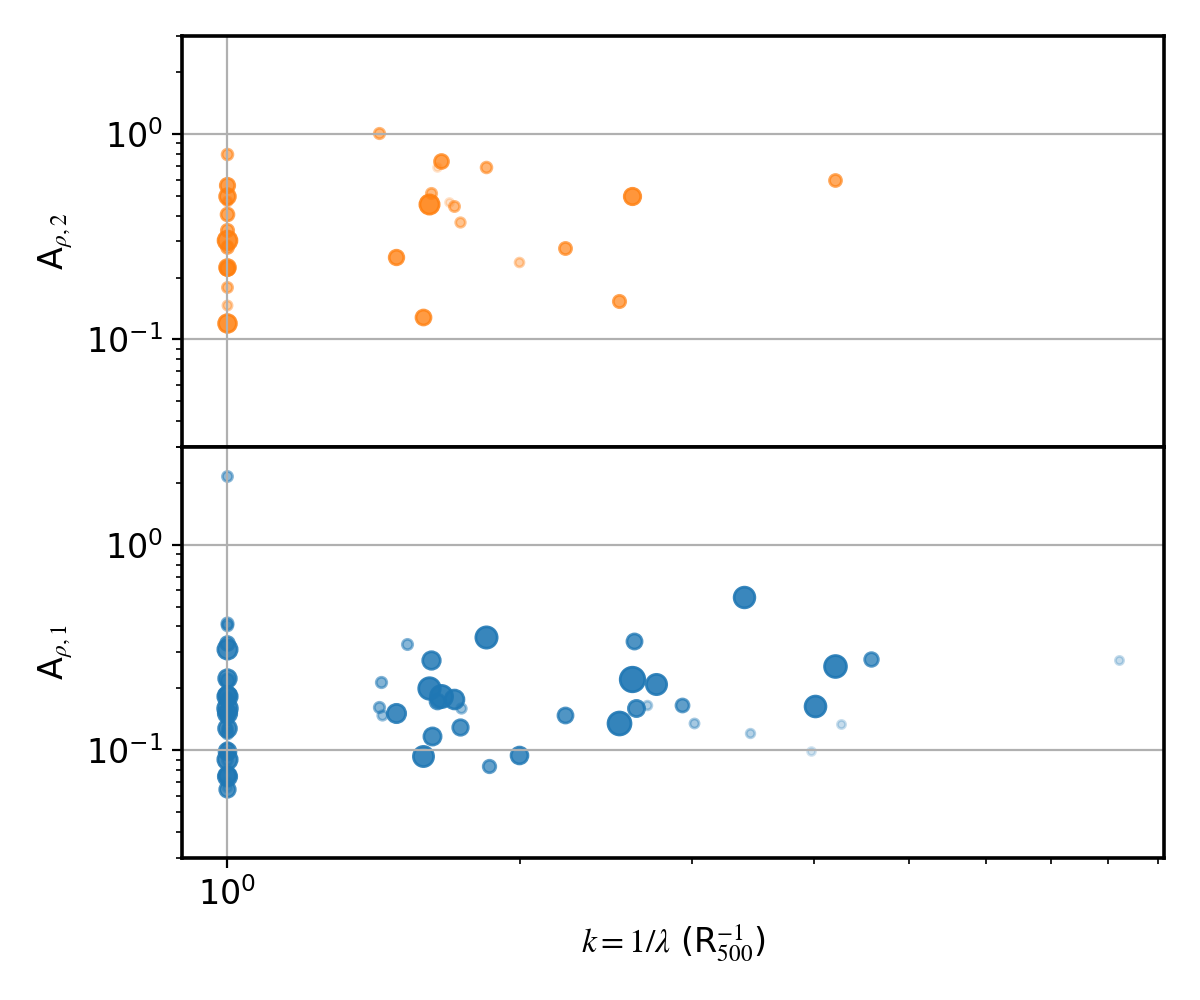}
        \caption{Nodes of amplitude spectra of density fluctuations where $A_{\rho}$ has significance $\xi > 2$ in the respective Ring. The size and transparency are scaled by the significance such that larger and more opaque points have greater statistical significance.}
        \label{fig:A3Ds_dots}
    \end{figure}

    We present all nodes of the amplitude spectra of density fluctuations, $A_{\rho}$, with significance $\xi > 2$ in Figure~\ref{fig:A3Ds_dots}. We see a clear trend of fewer points at higher $k$ (smaller scales), given the increased difficulty of placing constraints at these values (see Section~\ref{sec:toward}). Even so, we see in Ring 1 (bottom panel of Figure~\ref{fig:A3Ds_dots}) that there appears to be an upward trend in the amplitude spectra. This reflects the notion that the injection scales within Ring 1 are generally smaller than $R_{500}$. However, an average injection scale in either Ring is fairly unconstrained as we do not see a clear peak/turnover in the amplitude spectra.

%% file: Appendix_Ring2.tex
    We find 15 clusters for which at least one node in the amplitude spectra of density fluctuations within Ring 2 are at least $2\sigma$. The weighted average of the inferred Mach numbers is $\mathcal{M}_{\rho,2} = 0.87$, though the scatter is 0.89, where the distribution is asymmetric. Indeed, many of the systems have inferred gas velocities which are supersonic, which is not expected for turbulent motions alone. If, as before, we exclude those with inferred supersonic gas velocities, we arrive at only four clusters whose weighted average Mach number is $\mathcal{M}_{\rho,2} = 0.59 \pm 18$. 

    Of the 15 clusters, those which we infer to have supersonic gas velocities are: SPT-CLJ0658-5556, SPT-CLJ0638-5358, SPT-CLJ0438-5419, SPT-CLJ0304-4401, SPT-CLJ2023-5535, SPT-CLJ0114-4123, SPT-CLJ0014-3022, SPT-CLJ2341-5119, SPT-CLJ2146-4633, SPT-CLJ2032-5627, and SPT-CLJ0254-5857.
    Several of these are again known merging clusters where merging structure exists within Ring 2. 
    %We note that the surface brightness profile, $\bar{S}_{\text{ICM}}$, plays a role in both the numerator and denominator of the normalized residual images, $\delta S / \bar{S}_{\text{ICM}}$. Relative to the unmasked residuals, masking substructure may yield a residual image ($\delta S$) with smaller amplitudes. However, if a steeper slope has been fit (when masking substructure), then $\delta S / \bar{S}_{\text{ICM}}$ may in fact be larger than its unmasked counterpart. Another effect is that, for a fixed a $P_{\text{2D}}$, its deprojection will produce larger values of $P_{\text{3D}}$, and thus $A_{\text{3D}}$, relative to its counterpart from a surface brightness profile with a shallower slope. 
    As stated in Appendix~\ref{sec:substructure_masking}, masking substructure need not always reduce the inferred fluctuations as the masking can alter the fitted surface brightness profile.

    %We thus have the potential that we do see a veritable dichotomy between turbulent and non-turbulent motions within Ring 2, but we recognize that ability to confidently distinguish between them is reduced relative to Ring 1. 
    The distribution of inferred $\mathcal{M}_{\rho,2}$ appears bimodal as did the distribution of $\mathcal{M}_{\rho,1}$. This bimodality may have the same causation as in Ring 1, i.e. seeing turbulence versus substructure, where the latter is due to merging activity and likely corresponds to non-turbulent motions (e.g. shocks). However, we are unable to confidently assert the cause of the apparent bimodality in $\mathcal{M}_{\rho,2}$ due to the fainter X-ray signal in Ring 2.
    For various significance cuts, we have either three or four clusters with subsonic velocities, where the weighted means of $\mathcal{M}_{\text{3D}}$ are between 0.6 and 0.7, which is in agreement with expectations derived in Section~\ref{sec:discussion}. That said, better constraints over more clusters are clearly necessary to robustly distinguish between turbulent and non-turbulent motions out to $R_{500}$.
    
    %Thus, while $A_{\text{2D}}$ has uncertainties computed from the chains of the MCMC surface brightness profile fits, the window functions used for deprojection do not incorporate this uncertainty. 

    %The clusters for which $\mathcal{M}_{\text{3D}} > 1$ are: SPT-CLJ0354-5904, SPT-CLJ0658-5556, SPT-CLJ2017-6258, SPT-CLJ2056-5459, SPT-CLJ0304-4401, and SPT-CLJ0014-3022. Of these, SPT-CLJ0354-5904, SPT-CLJ2017-6258, and SPT-CLJ2056-5459 show evidence of disturbance; however, our substructure algorithm did not identify any substructure to mask in part due to the modest photon counts in those images. SPT-CLJ0658-5556 (the Bullet cluster), SPT-CLJ0014-3022 (Abell 2744), SPT-CLJ0304-4401, and SPT-CLJ2032-5627 are known mergers \citep{markevitch2002,kempner2004,raja2021,duchesne2021}. 